\documentclass[%
 aip,
 amsmath,amssymb,
 reprint,%
]{revtex4-1}

\usepackage{graphicx}
\usepackage{dcolumn}
\usepackage{bm}

\usepackage[utf8]{inputenc}
\usepackage[T1]{fontenc}
\usepackage{mathptmx}
\usepackage{siunitx}
\usepackage{etoolbox}
\usepackage{subcaption}
\usepackage{subfig}
\makeatletter
\def\@email#1#2{%
 \endgroup
 \patchcmd{\titleblock@produce}
  {\frontmatter@RRAPformat}
  {\frontmatter@RRAPformat{\produce@RRAP{*#1\href{mailto:#2}{#2}}}\frontmatter@RRAPformat}
  {}{}
}%
\makeatother
\begin{document}

\preprint{}

\title[Turbulent Mixing Dynamics of Under-Expanded Hydrogen Jets in Propulsion Systems]{Turbulent Mixing Dynamics of Under-Expanded Hydrogen Jets in Propulsion Systems}
\author{Francesco Duronio}
  \email[Corresponding author: ]{francesco.duronio@univaq.it}
 \altaffiliation[Also at ]{Consiglio Nazionale delle Ricerche, Istituto di Scienze e Tecnologie per l'Energia e la Mobilità Sostenibili (STEMS), Via G. Marconi 4, 80125 Napoli, Italy}
\author{Andrea Di Mascio}%
\affiliation{ 
Dipartimento di Ingegneria Industriale Informazione e di Economia - Università degli Studi dell'Aquila Piazzale Ernesto Pontieri, Monteluco di Roio, 67100, L'Aquila (AQ), Italy
}%

\date{\today}

\begin{abstract}
Underexpanded jets are present in various engineering applications; in recent years, they have gained special attention because of the development of gas-fueled propulsion systems. In these apparatuses, the direct injection of fuels such as hydrogen in innovative low-emission engines' chambers induces turbulent under-expanded jets. In this study, we performed high-fidelity Large Eddy Simulations of under-expanded hydrogen jets to investigate mixing characteristics and provide valuable insights for developing injectors suitable for hydrogen and, more generally, gaseous-fueled propulsion systems. We initially assessed the method's accuracy, evaluating the convergence and uncertainty of the numerical results and validating them against experimental particle image velocimetry and Schlieren data. The simulated jets, the Mach disc dimensions, and the resulting velocity field align closely with the experimental observations.  Then, we analysed the jet structure for pressure ratios of 4 to 25 and examined the effects of the geometrical configuration of the nozzle on the characteristics of the air-fuel mixture obtained.  We compared the jets resulting from a round-hole nozzle with annular ones resembling outward-opening injectors.  
\end{abstract}

\maketitle
\section{Introduction}
Under-expanded jets are complex high-speed flows in multiple engineering applications, such as aircraft engine exhaust plumes, rocket discharge, and combustion chamber injectors. Additionally, these jets can occur in natural phenomena, like volcanic eruptions \cite{https://doi.org/10.1029/2009JB006985,volcano2,gmd-6-1905-2013,doi:10.2514/3.49166}. 
Consequently, under-expanded jets have been extensively studied, particularly in aerospace applications. However, in the past decade, research has increasingly focused on their role in the injection process of advanced propulsion systems, positioning this as an emerging field within fluid dynamics and engine simulations \cite{en16186471,duronio2021pof}.
Although gas jets have been extensively studied for aerospace applications, it is equally crucial to develop a comprehensive understanding of these processes for propulsion systems \cite{doi:10.1177/14680874221081947}. 

In modern propulsion systems, fuels such as hydrogen, propane, ammonia, and methane are often injected as gases rather than liquids. 
Due to the substantial pressure difference between the injector rail and the injection environment, supersonic conditions are generally achieved \cite{en16186471,Vuorinen2014,Hamzehloo2014,Hamzehloo2016}. 
This leads to the formation of under-expanded jets downstream of the injector nozzles, resulting in a distinctive flow field structure \cite{Allocca2020,SAMSAMKHAYANI2022107144}. 
Among the different combustibles, hydrogen is gaining attention and research efforts. 
Direct injection into the combustion chamber is one of the most promising technologies that will likely be chosen for developing hydrogen combustion. 
In this context, it follows that the injection process plays a relevant role in the chain of events that takes place in the propulsion system. The injectors and the resulting under-expanded jets must be deeply investigated to obtain the desired air-fuel mixture, efficient combustion, and reduced tailpipe emissions.

Extensive research has been conducted on methane jets \cite{Yosri2020,Banholzer2019,BARTOLUCCI2020108660,Dong2018}, but comparatively less focus has been given to hydrogen fuel despite its growing scientific relevance. Schlieren imaging is commonly used to capture the evolution of transient under-expanded hydrogen jets from gaseous fuel injectors, evaluating factors like pressure ratio (PR), nozzle characteristics, and jet tip penetration \cite{CORATELLA2024432,YEGANEH2023125762,LEE20214538,2023-01-0320,2022-01-1009}. However, experimental studies on hydrogen jets are often limited to qualitative observations. In contrast, jets of other species are examined using density maps and Planar Laser Induced Fluorescence, providing local values of fuel concentration and gas density \cite{yu2013visualization,Yu2011,Sakellarakis2021,Ni2022}. Such detailed investigations are crucial for robust validation of new methods and CFD simulations of under-expanded jets.

Considering the intrinsic characteristics of the physical process under attention, simulation tools play a crucial role in gaining deeper insights into the physics of under-expanded jets and supporting the development of simplified models for designing and optimising injection devices. Common methodologies employ both explicit \cite{doi:10.1177/14680874221081947,Buttay2016,Hamzehloo2019,2022-01-0505} and implicit time integration algorithms \cite{Yosri2020,Rahantamialisoa2022,hamze2014}. Among these, using a Large Eddy Simulation (LES) turbulence framework and high-order integration schemes with low numerical dissipation has proven to be particularly effective for replicating the characteristics of under-expanded jets and the mixture formation \cite{HAMZEHLOO201421275,Vuorinen2014,Vuorinen2013}.

The literature covers extensive studies on under-expanded jets for gases like methane, nitrogen, and air \cite{Yu2013,Banholzer2017,Traxinger2018,Banholzer2019,Xiao2019}, with increasing recent interest in hydrogen \cite{Anaclerio2022,anaclerio_capurso_torresi_2023,BALLATORE2024771,10.1063/1.5144558}.
Hamzehloo et al. \cite{Hamzehloo2016b,Hamzehloo2014,Hamzehloo2016} explored near-nozzle characteristics of various under-expanded jets using a CFD code based on the Advection Upstream Splitting Method (AUSM) discretization for compressible flows, focusing on jet tip penetration, shear layers, and Mach disc structures.
Cryogenic hydrogen (H$_2$) jets were also experimentally and numerically investigated to examine the nozzle diameter and pressure ratio effects on jet expansion structures \cite{Ren2020, HECHT20198960}. The numerical setup of Ren et al. \cite{Ren2020} used a 2D model with WENO schemes to optimise computational efficiency; their simulation differs from other typical setups because it uses a total pressure boundary condition instead of a high-pressure reservoir; the same approach was also adopted by Zhang et al. \cite{zhang_aubry_chen_wu_sha_2019}.
Further research has examined how real-fluid properties influence jet behaviour. Studies comparing the Redlich-Kwong and Peng-Robinson equations of state revealed that, under certain injection conditions, results differ significantly from those based on the ideal-gas law \cite{Jin2021,Rahantamialisoa2023}. In particular, adopting the real fluid model, the Mach number results in higher values within the first shock-cell, while temperature achieves lower values downstream of the Mach disc using the ideal gas equation. This last difference, in turn, affects the mass flow predicted.
However, the most important limitation of all these studies is the computational resources required. Historically, injection processes have been studied relying on Eulerian-Lagrangian CFD codes because they involve liquid fuels. So, the minimum size of the grid is approximately equal to the diameter of the nozzle \cite{DURONIO2025105048,DURONIO2025100991}. 
In contrast, with gaseous fuels, a correct grid size, capable of correctly representing the under-expanded jets, is of the order of magnitude of $D/(30\div50)$ \cite{HAMZEHLOO201421275}, where D is the nozzle diameter. This completely changes the requirements regarding resources needed and poses important limitations on the simulation's feasibility. Different meshing strategies have been adopted to reduce the computational load. However, the common approach is to use multiple refinement regions with the grid size gradually increasing downstream of the exit section of the nozzle \cite{2022-01-0505,Hamzehloo2016b}. This allowed researchers to perform the simulations in a reasonable time, but reduced the quality of the results, not correctly predicting the characteristics of the air/fuel mixture, especially when using RANS turbulence models \cite{10.1115/ICEF2024-140413}.
Other studies even simplify the problem of running 2D simulations \cite{Anaclerio2022,Ren2020,zhang_aubry_chen_wu_sha_2019}.
All these approaches are unreliable for developing propulsion systems and also for the simulation of the complete engine cycle \cite{BALLATORE2024771,LUAN2025105535,en16186471}.
The present paper shows a CFD investigation of hydrogen under-expanded jets related to propulsion applications. Unlike in the past, we studied these hydrogen jets focusing on turbulent mixing; the investigation provides information for the optimal design of injection devices suitable for gaseous-fuelled propulsion systems.
We ran high-fidelity GPU-accelerated simulations adopting high-resolution grid sizing for the jet volume. We initially validate our simulations by relying on quantitative Particle Image Velocimetry (PIV) and Schlieren images of under-expanded jets; then, we investigate different injection configurations with different pressure ratios and evaluate the jet's structure as well as the mixture formation process.
We also analyse the nozzle characteristics, showing how the flow drastically changes with annular nozzles and hollow cone jets, enhancing the mixing process.

\section{Mathematical Models and Numerical Methods}
\subsection{Mathematical Models}
The numerical simulation of the compressible multi-species flow is performed by the integration of the following governing equations:
\begin{equation}
\left\{
\begin{array}{l}
\displaystyle
 \frac{\partial \rho}{\partial t} + \nabla \cdot(\rho \mathbf{u})= 0 \\*[0.5cm]
\displaystyle
 \frac{\partial(\rho \mathbf{u})}{\partial t} +\nabla \cdot(\rho \mathbf{u} \otimes \mathbf{u})= 
 -\nabla p +\nabla \cdot \boldsymbol{\Pi} \\*[0.5cm]
\displaystyle
 \frac{\partial(\rho E)}{\partial t} +\nabla \cdot(\rho \mathbf{u} E)=
 \nabla \cdot(\boldsymbol{\Pi} \cdot \mathbf{u} - p \mathbf{u})+\nabla \cdot \boldsymbol{\mathcal{Q}}  \\*[0.5cm]
\displaystyle
 \frac{\partial\left(\rho Y_k\right)}{\partial t}+\nabla \cdot\left(\rho \mathbf{u} Y_k\right)=
 \quad \nabla \cdot \boldsymbol{\mathcal{F}_k} 
 \end{array}
\right.
\label{GovEq}
\end{equation}

In the above equations, $\rho, \mathbf{u}$, and $p$ are the density, velocity vector, and pressure, respectively. $E=e+\mathbf{u} \cdot \mathbf{u} / 2$ is the specific total energy with $e$ representing the specific internal energy; $Y_k$ is the mass fraction of the $k^{th}$ species.
The viscous stress tensor, $\boldsymbol{\Pi}$, under the Newtonian assumption, is given by:
\begin{equation}
\boldsymbol{\Pi} = \mu \left[\nabla \mathbf{u}+(\nabla \mathbf{u})^{\mathrm{T}}\right] +  \lambda  \mathbf{I} \left( \nabla \cdot \mathbf{u}\right)
\end{equation}
where $\mu$ and $\lambda$ are dynamic viscosity and the Lam\`e constant ($\lambda = - 2/3\mu$), respectively.
$\boldsymbol{\mathcal{Q}}$ is conduction heat flux:
\begin{equation}
    \boldsymbol{\mathcal{Q}} = \kappa\nabla T
\end{equation}
where $\kappa$ is the thermal conductivity. 
$\boldsymbol{\mathcal{F}_k}$ is the diffusive transport flux of the k-th species:
\begin{equation}
    \boldsymbol{\mathcal{F}_k} = \rho Y_k D_{k,j} \nabla Y_k
\end{equation}
where $D_{k,j}$ is the binary diffusivity coefficient between the species. 
We modeled the fluid with the perfect gas equation of state.
In all the simulations shown in the paper, we adopted the Dynamic Smagorinsky LES model \cite{10.1063/1.857955,smago} to accurately resolve the turbulent structures of the jet.
This model has already been used in the past to simulate under-expanded jets \cite{doi:10.2514/1.J051470,doi:10.2514/1.J050362,doi:10.2514/1.J054689,buttay2016analysis}.
For the sake of brevity, we are not reporting full details; the reader is referred to the original papers for a complete description.
We set the Schmidt and Prandtl numbers equal to 0.7, following previous works \cite{Vuorinen2014,Hamzehloo2019,2022-01-0505}.

\subsection{Numerical Method}
The numerical integration of the governing equation \eqref{GovEq} is performed using the AmRex PeleC solver\cite{PeleSoftware}, which supports block-structured adaptive mesh refinement (AMR) and GPU parallelisation \cite{Zhang2019pele}. In the code, the inviscid fluxes in \eqref{GovEq} are discretised using the unsplit piecewise parabolic method (PPM) with hybrid PPM WENO variants \cite{COLELLA1984174,motheau2020investigation}.
The WENO reconstruction is performed with the 7th-order WENO-Z scheme of \cite{BALSARA2000405}.
The diffusion fluxes are discretised in space with second-order centred differences. 
Temporal integration relies on a standard predictor-corrector approach \cite{PeleSoftware,PeleC_IJHPCA}.
Transport coefficients are evaluated from the CHEMKIN transport library and depend on temperature \cite{kee1986fortran,ERN1995105}.
Thermodynamic properties are evaluated using the NASA7 polynomial parametrisation \cite{mcbride2002nasa}.

\section{Test cases: Characteristics and Numerical Setup \label{sec:testcases}}
We investigated seven different case studies covering different morphologies of the under-expanded jets. Table \ref{tab:cases} reports their main characteristics.
\begin{table*}
\caption{\label{tab:cases}Test cases details. $PR$: pressure ratio; $p_{inj}$: injection pressure; $p_{amb}$: ambient pressure; 
           $T_{amb}$: ambient temperature; $D$: nozzle diameter; $(^*):$ External diameter.}
\begin{ruledtabular}
\begin{tabular}{cccccccccc}
        Case N$^\circ$&  Nozzle & Gas/amb  & $PR$  &  $p_{inj}$ [bar]&  $p_{amb}$ [bar] & $T_{amb}$ [K] & $D [mm]$ & Jet cone angle [$^\circ$] \\[0.1cm]
        \hline
         Validation 1 &Round  & H$_2$/Air&  10  & 9.976 & 0.9976 & 292 & 1.5 & N/A   \\
         Validation 2 &Round  & Air/Air & 4.2  & 4.25 & 1 &  288 & 15 & N/A   \\
         R1 &Round  & H$_2$/Air&  4.2  & 4.25 & 1 &  288 & 15& N/A   \\
         R2 &Round  & H$_2$/Air&  12.6   & 12.8 & 1&  288 & 15 & N/A  \\
         R3 &Round & H$_2$/Air&  25.2 &25.6 & 1 &  288  & 15 & N/A \\
         A4 &Annular  & H$_2$/Air&  12.6 & 12.8 & 1 &  288 & 15$\sqrt{2}$ $(^*)$ & 90  \\
         A5 &Annular & H$_2$/Air&  12.6  & 12.8 & 1&  288 & 15$\sqrt{2}$ $(^*)$ & 135 \\
\end{tabular}
\end{ruledtabular}
\end{table*}

\begin{figure}[h!]
\begin{subfigure}{.6\columnwidth}
  \centerline{\includegraphics[width=\linewidth]{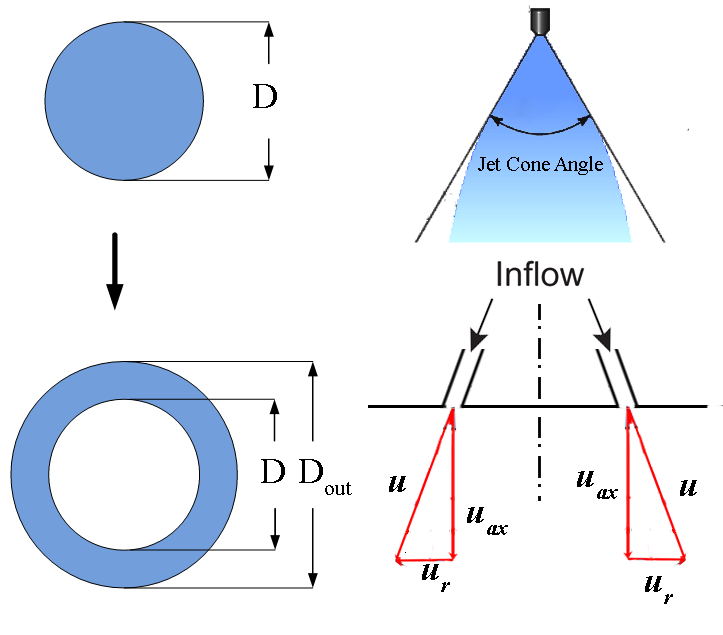}}
    \caption{}
    \label{fig:nozzles}
\end{subfigure}
\begin{subfigure}{.6\columnwidth}
  \centerline{\includegraphics[width=\linewidth]{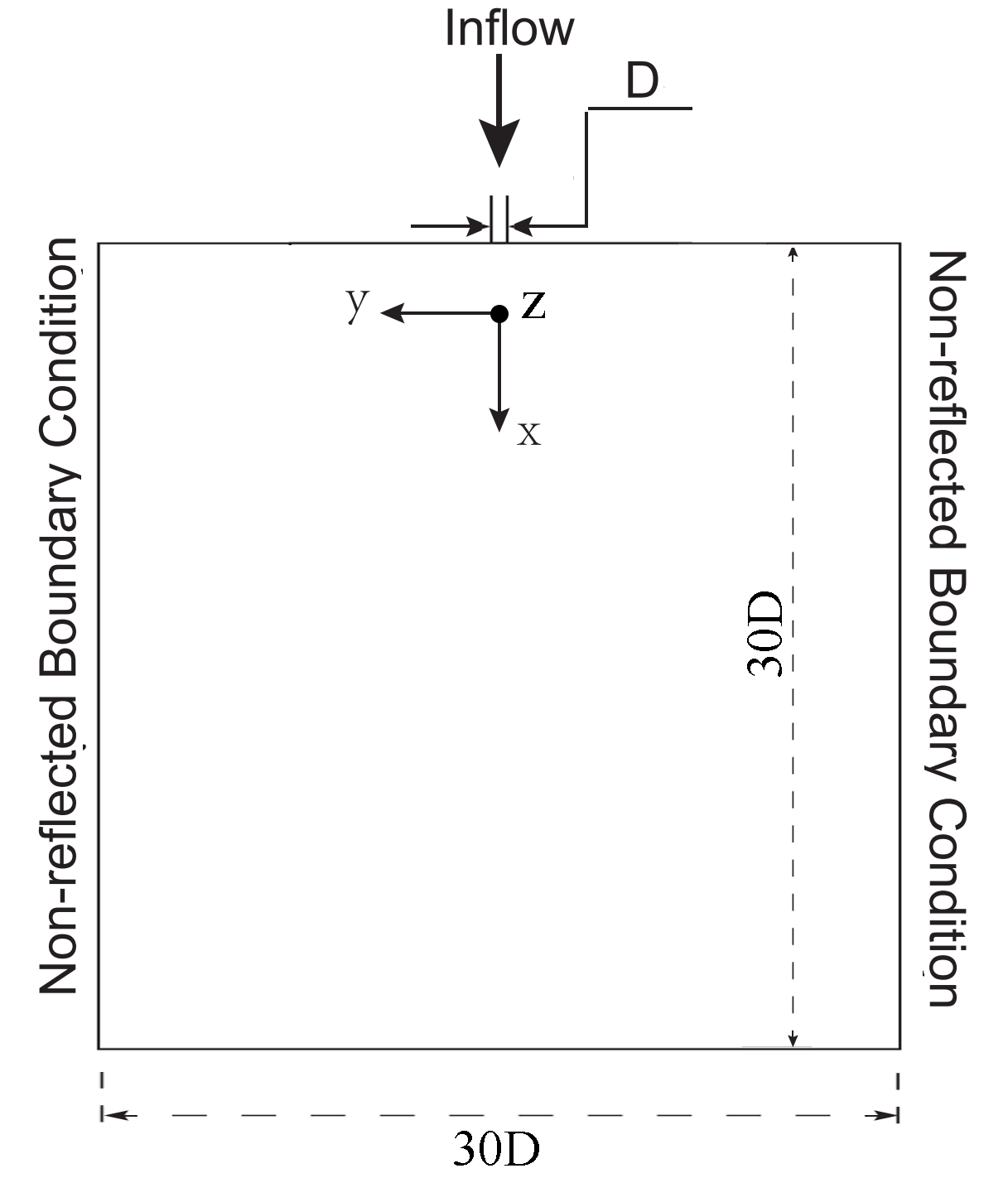}}
    \caption{}
    \label{fig:domain}
\end{subfigure}
  \caption{ Nozzle configuration investigated and jet cone angle definition Figure \ref{fig:nozzles}. Overview of the computational domain Figure \ref{fig:domain}}
\label{fig:domainGrid}
\end{figure}

For verification and validation of the simulations, we considered the jets experimentally investigated by \citet{RUGGLES201217549} and \citet{10.1063/1.4894741}. Both are round jets issued from circular nozzles. The diameter is equal to \SI{1.5}{\milli\meter} for nozzle 1 and \SI{15}{\milli\meter} for nozzle 2.
After these first verification cases, we performed other simulations in which we replicated case 2 by replacing air with hydrogen and then varied the pressure ratio $PR$ from 4.2 to 25. The last two test cases A4 and A5 deal with annular nozzles, as shown in Figure \ref{fig:nozzles}. We defined them by maintaining the same inner diameter of cases R1-R3 and computing the outer diameter by imposing the same nozzle exit area. These configurations represent hollow cone injectors where the velocity of the gaseous fuel exhibits a radial component ($\mathbf{u_r}$), drastically changing the mixing process with the surrounding air. We studied two different values of the cone angle, representative of real prototypal devices \cite{2024-01-2617}.

We implemented the discrete equations on the 3D computational domain reported in Figure \ref{fig:domain}.

The injection environment is a cubic box with sides equal to $30D$. Table \ref{tab:cases} summarises inflow and initial conditions. Validation cases 1 and 2 replicate the test conditions of \citet{RUGGLES201217549} and \citet{10.1063/1.4894741}. We also imposed non-reflecting boundary conditions at the sides of the box, while on the bottom we enforced a zero-gradient boundary condition. We refined the mesh using fixed region refinements and Adaptive Mesh Refinement (AMR). In each region, the grid size obeys the law:
\begin{equation}
\Delta_{i}=\frac{\Delta_b}{2^{R_{L}}},
\end{equation}
where  $\Delta_i$ is the dimension of the grid in the generic refinement region, $\Delta_b$ is the base grid dimension, and $R_L$ is the refinement level. In all the simulations reported, we used $R_L=6$ and $\Delta_b=D$; the size gradually decreases from the maximum to the minimum dimension (placed around the nozzle) $\Delta_6=D/64$ with intermediate buffer levels. The region of maximum refinement around the nozzle has size $8D \mbox{(jet axis)}\times 3D \times 3D$.

We activated grid adaptation to guarantee the finest discretisation of the whole jet and turned it on when one of the following conditions was verified:  
\begin{enumerate}
    \item the velocity magnitude exceeds a threshold value $V_{thr}=\SI{3.5}{\meter\per\second}$, i.e.,$\mathbf{|u|}_{i,j,k}\geq V_{thr}$;
      \item the maximum difference of the density in adjacent locations exceeds a threshold value equal to $10^{-1} Kg \,\, m^{-3}$, i.e.
    \begin{equation}
\begin{array}{llll}
\max ( & \left|\rho_{i+1, j, k}-\rho_{i, j, k}\right|, & \left|\rho_{i, j, k}-\rho_{i-1, j, k}\right|, & \\*[3mm]
       & \left|\rho_{i, j+1, k}-\rho_{i, j, k}\right|, & \left|\rho_{i, j, k}-\rho_{i, j-1, k}\right|, & \\*[3mm]
       & \left|\rho_{i, j, k+1}-\rho_{i, j, k}\right|, & \left|\rho_{i, j, k}-\rho_{i, j,k-1}\right| & ) \geq 
        10^{-1} Kg \, \, m^{-3}
\end{array}
\end{equation}
\end{enumerate}

The integration interval depends on the time the jet takes to reach the domain bottom, which is $O(80 \div 240) D/U_{exit}$. 
As also discussed in various previous papers \cite{Vuorinen2013,Hamzehloo2019}, this duration guarantees that the structures and dimensions in the main jet body (width and height, shock positions, and so on ) oscillate around fixed values. We then computed the running averages by sampling the simulation's data after the onset of this almost steady-state phase. 
Finally, we built two coarser meshes for grid-dependence verification by doubling $\Delta_b$.

\section{Verification and Validation}

We assessed the convergence and uncertainty of the simulations using three grid levels, obtained as described at the end of section \ref{sec:testcases}.
The number of points in the grid for the finest level at the start of the simulation is around 12 million, while the maximum number of points during dynamic refinement is around 187 million.  
We followed the procedure outlined in the classical paper by \citet{roache1997quantification}, in which the procedures now adopted by AIAA, ITTC, and IEEE are described and discussed in detail. 

Table \ref{tab:unc} reports the average values for the main fields evaluated along the jet's axis and the relative Grid Convergence Index (GCI) based on $L_2$ norms of the field variations, evaluated in the whole field at the end of the simulations.
\begin{table}[h]
\caption{\label{tab:unc}Average values on the jet axis and grid convergence index evaluation for validation cases 1 and 2. }
\begin{ruledtabular}
\begin{tabular}{ccccc}
  &   \multicolumn{2}{c}{ Case 1}  & \multicolumn{2}{c}{ Case 2}    \\ \hline
&Average Value &  GCI  &Average Value&  GCI \\ \hline
 $\rho/\rho_{exit}$ & 0.25 & 1.54\% &    0.47&   1.47\%   \\
$p/p_{exit}$  & 0.19 &    1.27\% &  0.43 &   1.27\% \\ 
 $U_{x}/U_{exit}$ & 1.10  & 2.52\% & 1.17  &    2.84\%\\ 
\end{tabular}
\end{ruledtabular}
\end{table}

\begin{figure}[h!]
\centering
\includegraphics[width=.9\linewidth]{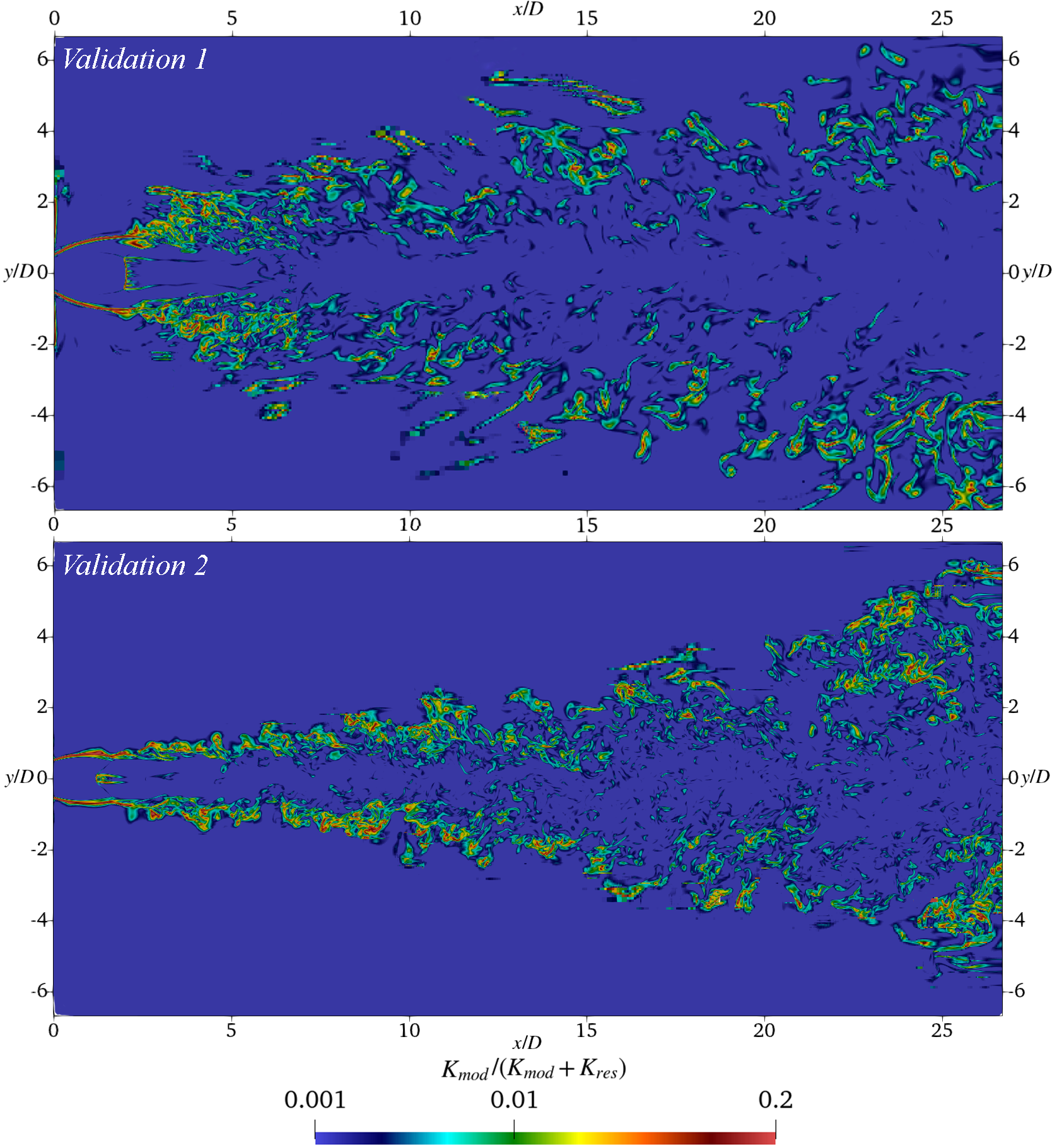}
\caption{ $K_{mod}/(K_{mod}+K_{res})$ plotted on an axial section of validation cases 1 and 2.}
\label{fig:K}
\end{figure}

To assess the adequacy of the chosen grid for Large Eddy Simulation, we evaluated the modelled kinetic energy and compared it to the total kinetic energy. This evaluation follows the methodology outlined by \citet {dimascio_dubbioso_muscari_2022}.
Figure \ref{fig:K} illustrates the instantaneous ratio between the modelled and total kinetic energy computed once the jet reaches statistically quasi-steady conditions on the axial midplane.

The model kinetic energy is significantly lower than the resolved one, and the modelled-to-total kinetic energy ratio is less than 0.2 for most of the domain (the highest values appear in the shear layer). The ratio, therefore, is below the limit of 0.3, which guarantees adequate grid resolution for Large-Eddy Simulation \cite{pope2000turbulent}.

\begin{figure}[h!]
	\begin{subfigure}[b]{.7\columnwidth}
\includegraphics[width=\linewidth]{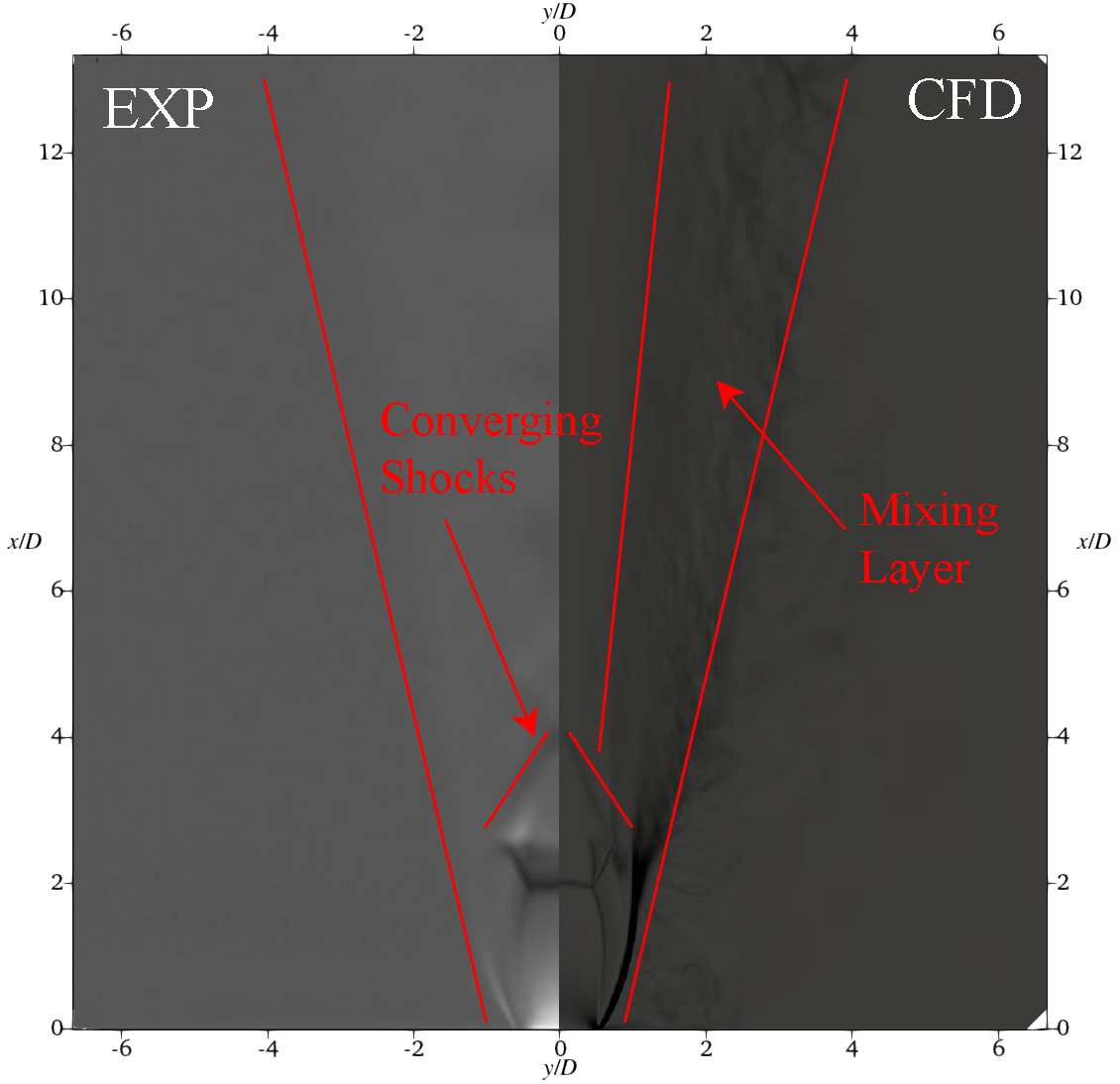}
	\end{subfigure}
	\begin{subfigure}[b]{.7\columnwidth}
\includegraphics[width=\linewidth]{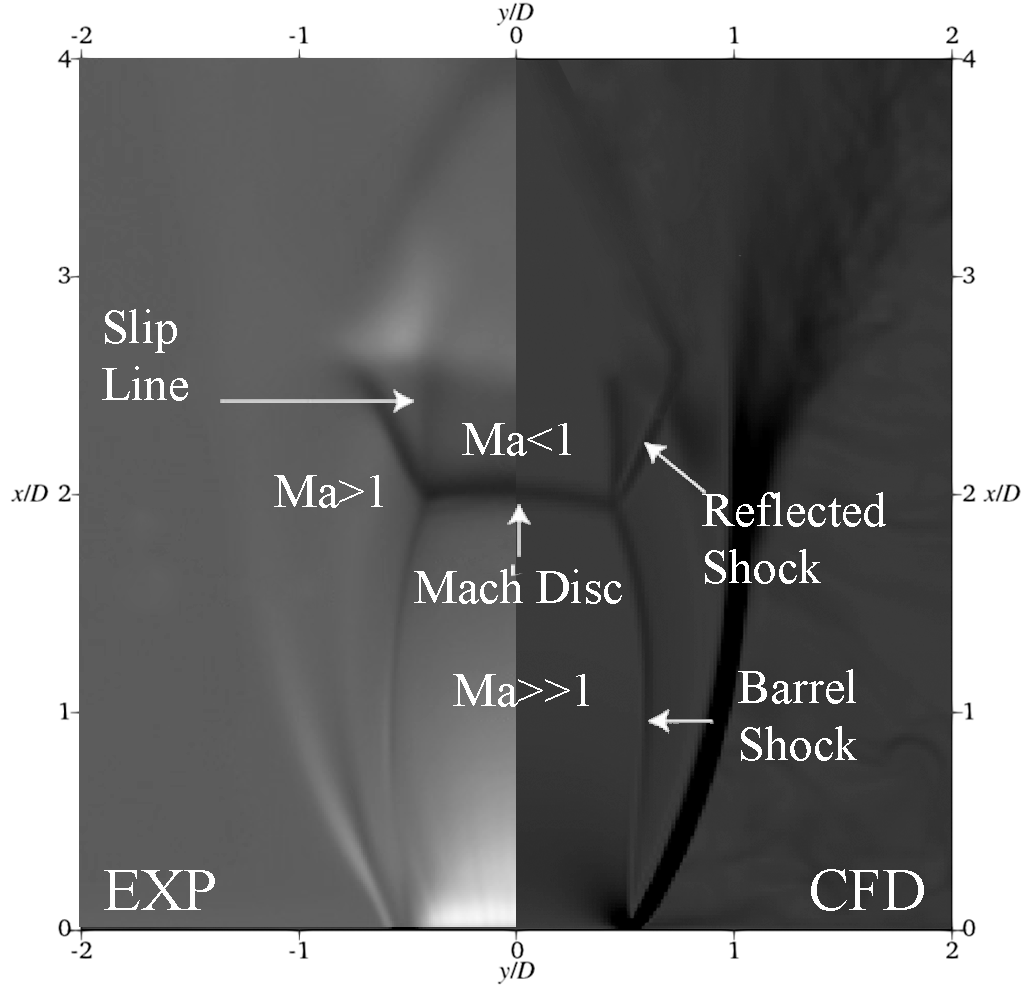}
	\end{subfigure}\caption{Validation case 1. Comparison of an elaboration of the Schlieren images acquired by \citet{RUGGLES201217549} and the CFD simulation performed in the current work. Reproduced with permission from Int. J. of Hydrogen Energy. 37, 22 (2012). Copyright 2012 Elsevier.}
\label{fig:compMorp}
\end{figure}

Next, we compared the numerical results with experimental data. Figure \ref{fig:compMorp} shows the time-averaged shape of the hydrogen jet and the Mach disc recorded using Schlieren imaging by \citet{RUGGLES201217549} and the numerical results in terms of time-averaged density gradient for validation case 1. 

As can be observed, the numerical simulation captures both the morphology of the hydrogen jet and the dimensions of the Mach disc. The jet has a peculiar conical shape bounded by a relevant mixing layer. 

\begin{figure}[h!]
\centering
\includegraphics[width=.95\linewidth]{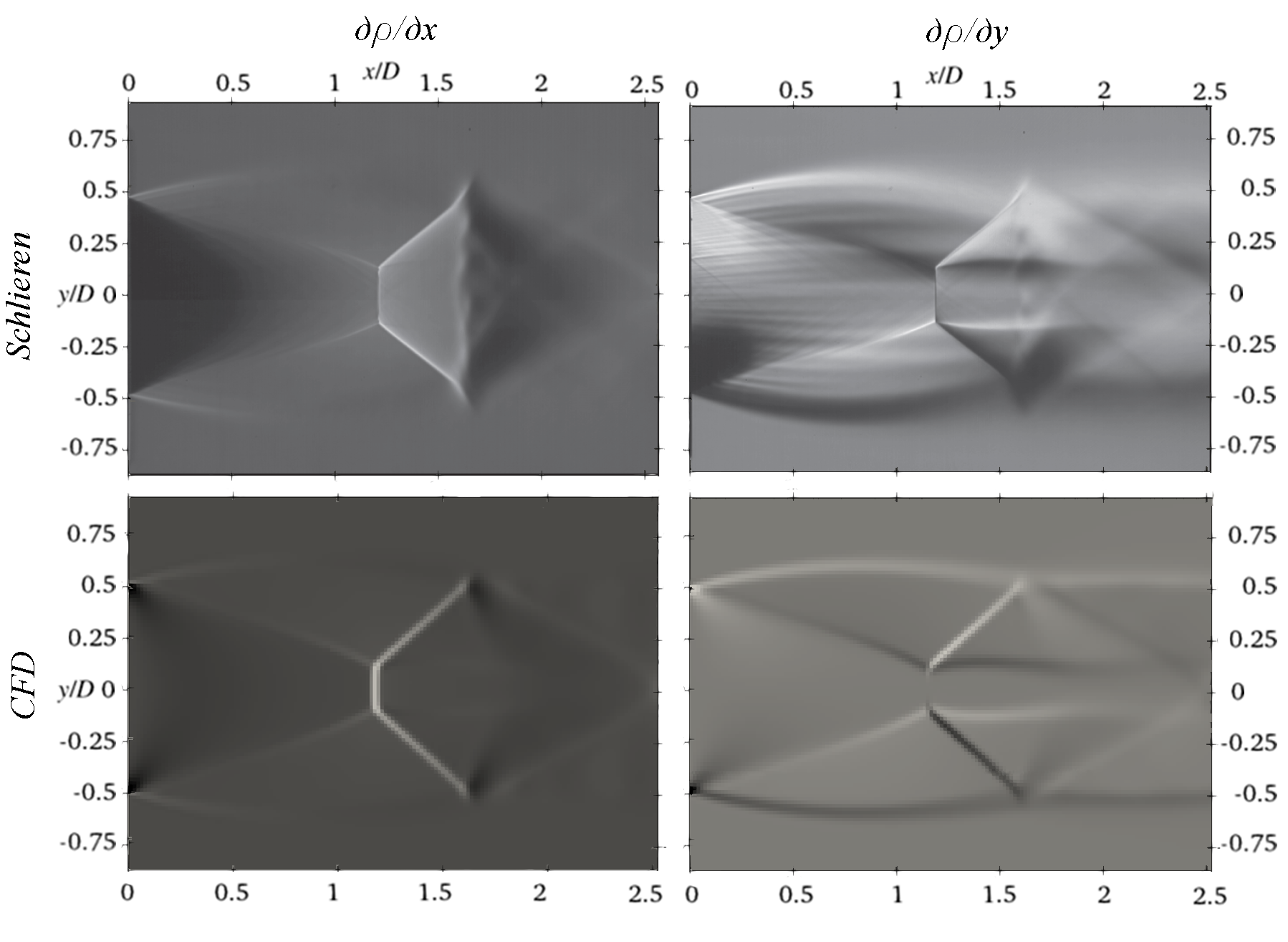}
\caption{ Validation case 2. Comparison of $\partial \rho/\partial x$ and $\partial \rho/\partial y$ from \cite{10.1063/1.4894741} and the present numerical simulation. Reproduced from Daniel Edgington-Mitchell, Damon R. Honnery, Julio Soria; The underexpanded jet Mach disk and its associated shear layer. Physics of Fluids 1 September 2014; 26 (9): 096101, with the permission of AIP Publishing.}
\label{fig:compMorp2}
\end{figure}

Figure \ref{fig:compMorp2} compares the partial derivatives for the density $\partial \rho/ \partial x$ and $\partial \rho/\partial y$, relative to validation case 2.

The overall agreement is good, and small differences are present only near the jet boundary in the oblique shocks downstream of the Mach disc.  
The jet is highly under-expanded, being the  $4<PR<7$ \cite{franquet2015free}. It has a “barrel” or “bottle” structure, and Mach disc appears (due to a singular reflection). The regular reflection of the intercepting shock on the axis is no longer possible. As a result, this reflection becomes singular, resulting in the appearance of a normal shock-denominated Mach disc.
The oblique shocks, slip-lines, the normal shock, and Prandtl-Mayer expansion fan features are well represented. As discussed in the paper by Edgington et al. \cite{10.1063/1.4894741}, the white stripes that appear in the shear layer of the experimental images are representative of zones of high standard deviation and so fluctuations of $\partial \rho/\partial y$.

\begin{figure}[h!]
\includegraphics[width=\linewidth]{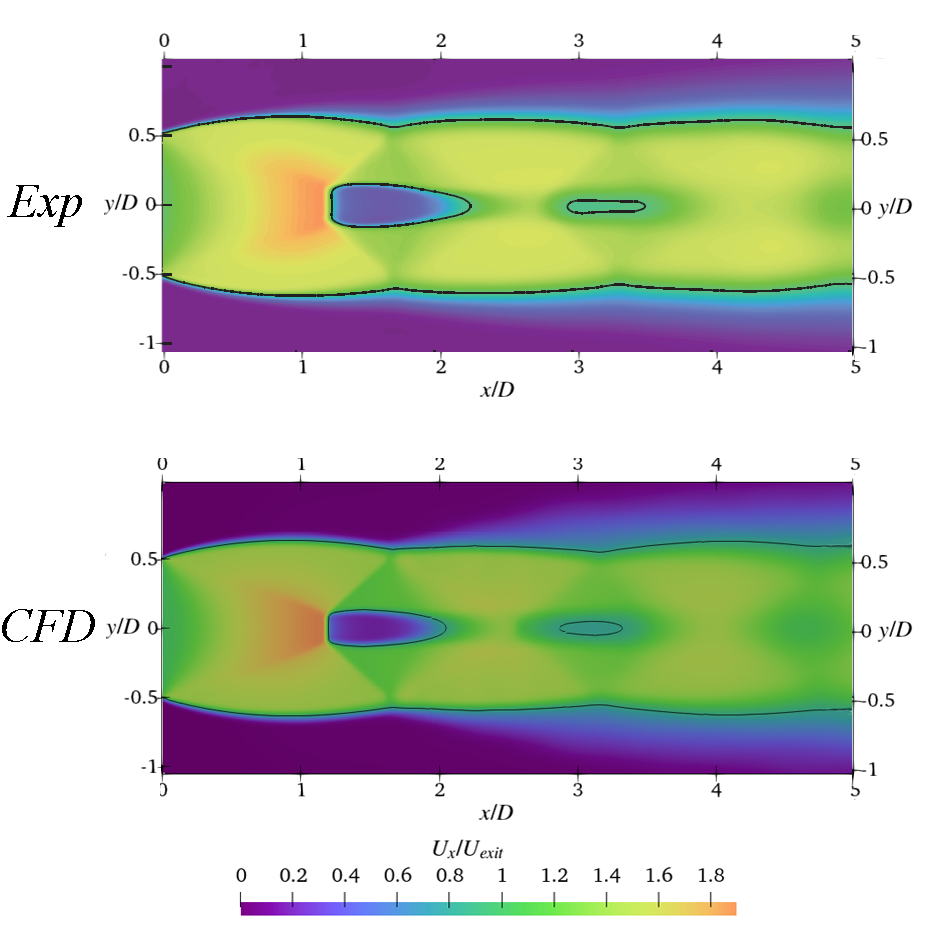}
\caption{Validation case 2. Comparison of experimental\cite{10.1063/1.4894741} and numerical non-dimensional axial velocities ($U_x$) colour maps. $U_{exit}$ is the speed of sound computed at the nozzle exit section thermodynamic conditions. Reproduced from Daniel Edgington-Mitchell, Damon R. Honnery, Julio Soria; The underexpanded jet Mach disk and its associated shear layer. Physics of Fluids 1 September 2014; 26 (9): 096101, with the permission of AIP Publishing.}
\label{fig:comp1}
\end{figure}

\begin{figure}[h!]
\includegraphics[width=\linewidth]{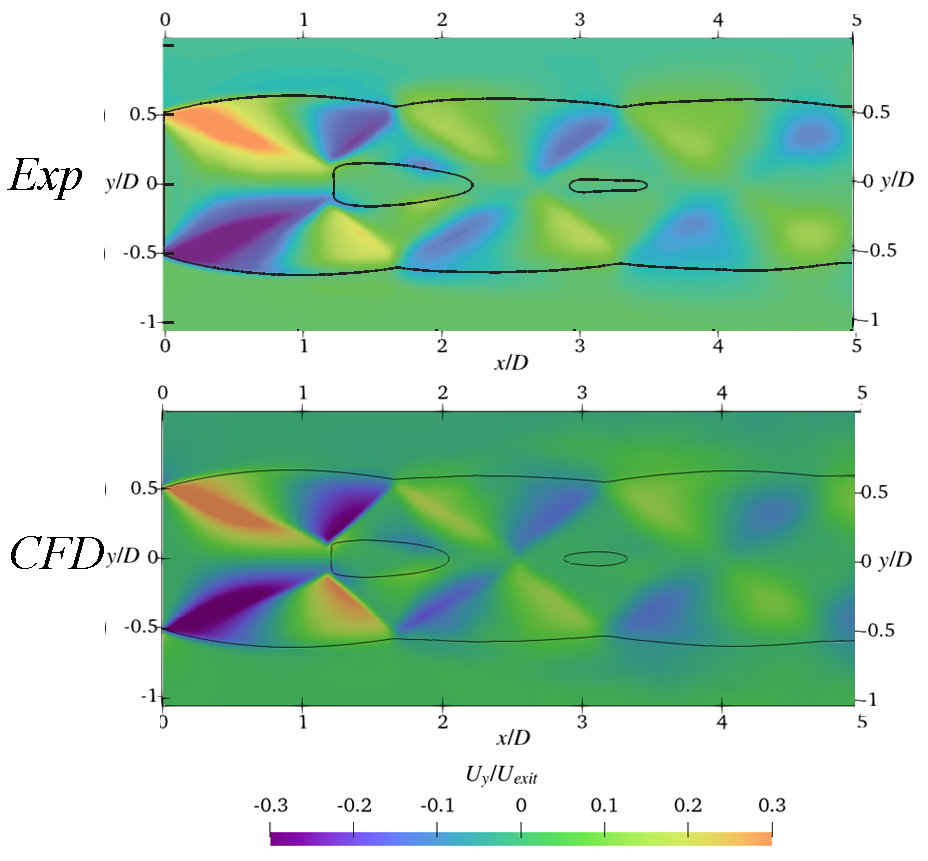}
\caption{Validation case 2. Comparison of experimental\cite{10.1063/1.4894741}  and numerical non-dimensional transverse ($U_y$) velocities colour maps. $U_{exit}$ is the speed of sound computed at the nozzle exit section thermodynamic conditions. Reproduced from Daniel Edgington-Mitchell, Damon R. Honnery, Julio Soria; The underexpanded jet Mach disk and its associated shear layer. Physics of Fluids 1 September 2014; 26 (9): 096101, with the permission of AIP Publishing.}
\label{fig:comp1_2}
\end{figure}

Figures \ref{fig:comp1} and \ref{fig:comp1_2} report the contours of mean axial and transverse flow velocities measured by \citet{10.1063/1.4894741} and computed in the present paper. 

We normalised all velocities with the value at the nozzle exit in the hypothesis of sonic conditions at the exit.
The simulations correctly reproduce the morphology of the velocity field for both the axial and transverse components. Moreover, the computed Mach disc matches the PIV data, and the magnitude of the velocity is correctly estimated.

\begin{figure}[h!]
	\begin{subfigure}[b]{\columnwidth}
\includegraphics[width=.65\linewidth]{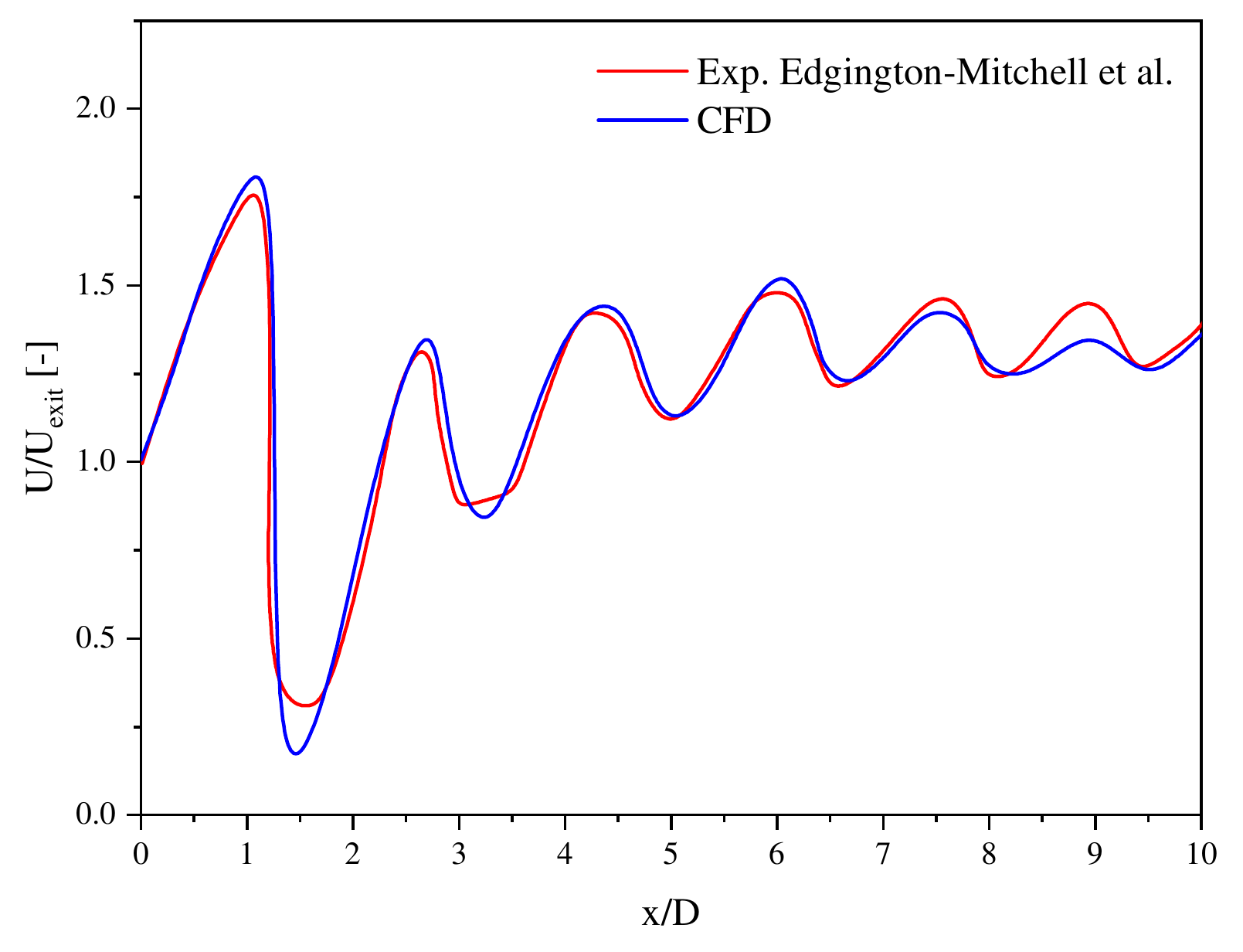}
	\end{subfigure}
	\begin{subfigure}[b]{\columnwidth}
\includegraphics[width=.65\linewidth]{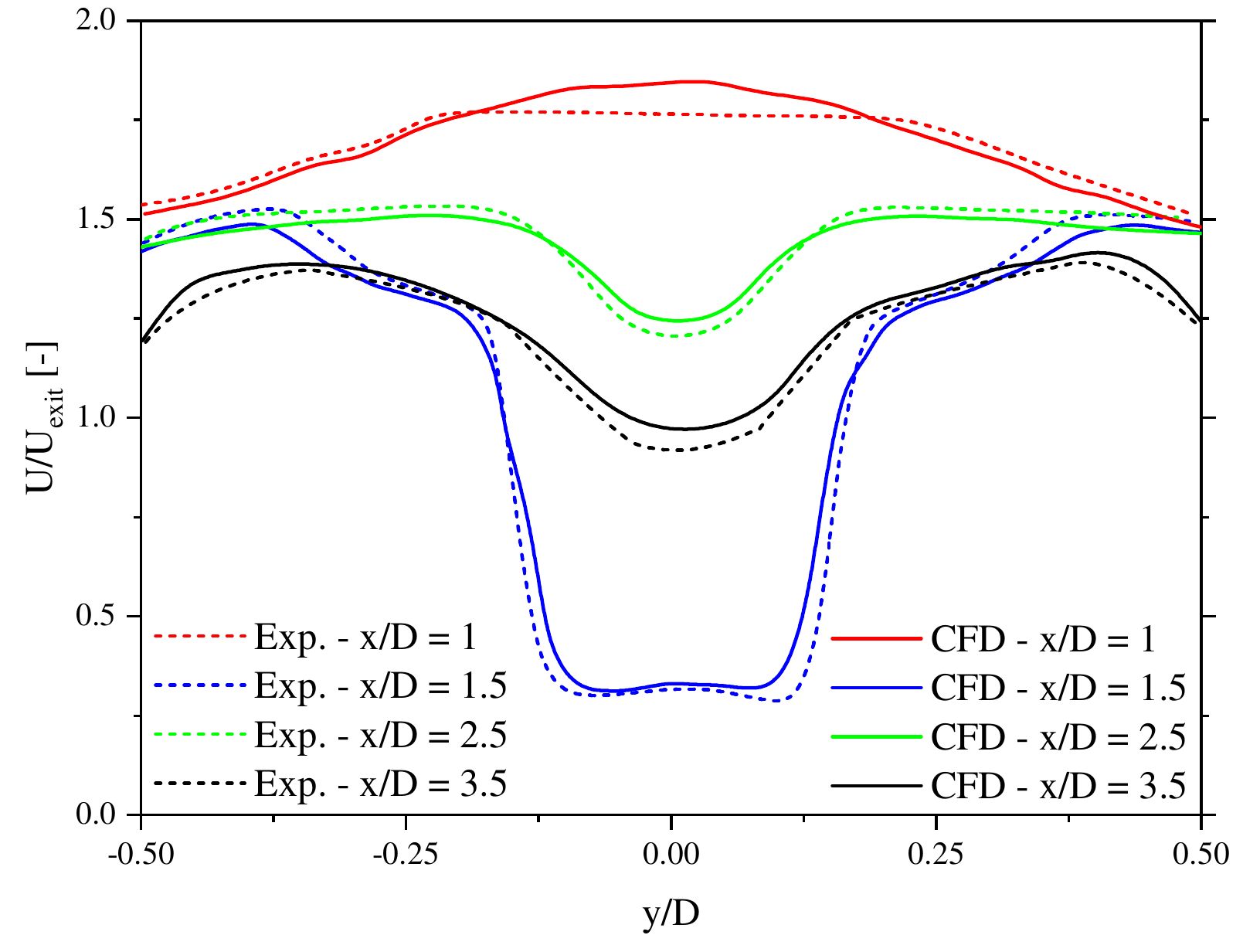}
	\end{subfigure}\caption{Validation case 2. Time-averaged axial velocity profiles. CFD results and experiments performed by \citet{10.1063/1.4894741}.}
\label{fig:comp2}
\end{figure}

Figure \ref{fig:comp2} compares the time-averaged velocity profiles along the jet axis and several transverse sections. The agreement between the experimental data and the simulation is satisfactory. 

\section{Discussion of the results}
We split the discussion into three parts. The first compares H${_2}$ and air under-expanded jets. Then, we investigate the pressure ratios' effects on the hydrogen jets' mixing process with the surrounding air. Finally, we examine the impact of the nozzle design (test cases A4 and A5) by varying the jet angle from 90$^\circ$ to 135$^\circ$.

\subsection{Under-expanded H${_2}$ and air jets}
\begin{figure}[h]
\includegraphics[width=.8\linewidth]{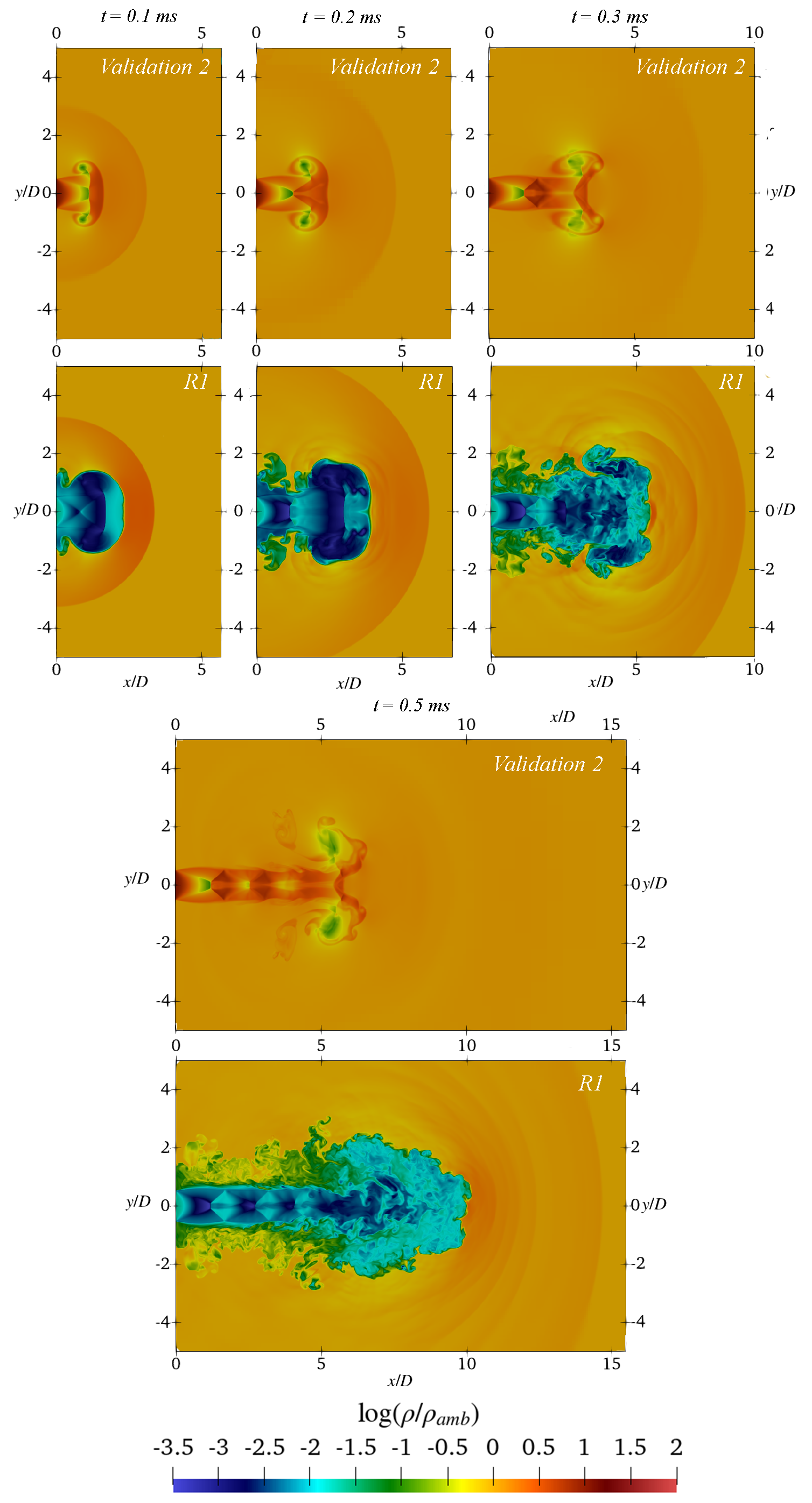}
\caption{Transient evolution of the jets of the validation case 2 and R1. Early stages.}
\label{fig:evCompAir}
\end{figure}

\begin{figure}[h]
\includegraphics[width=.8\linewidth]{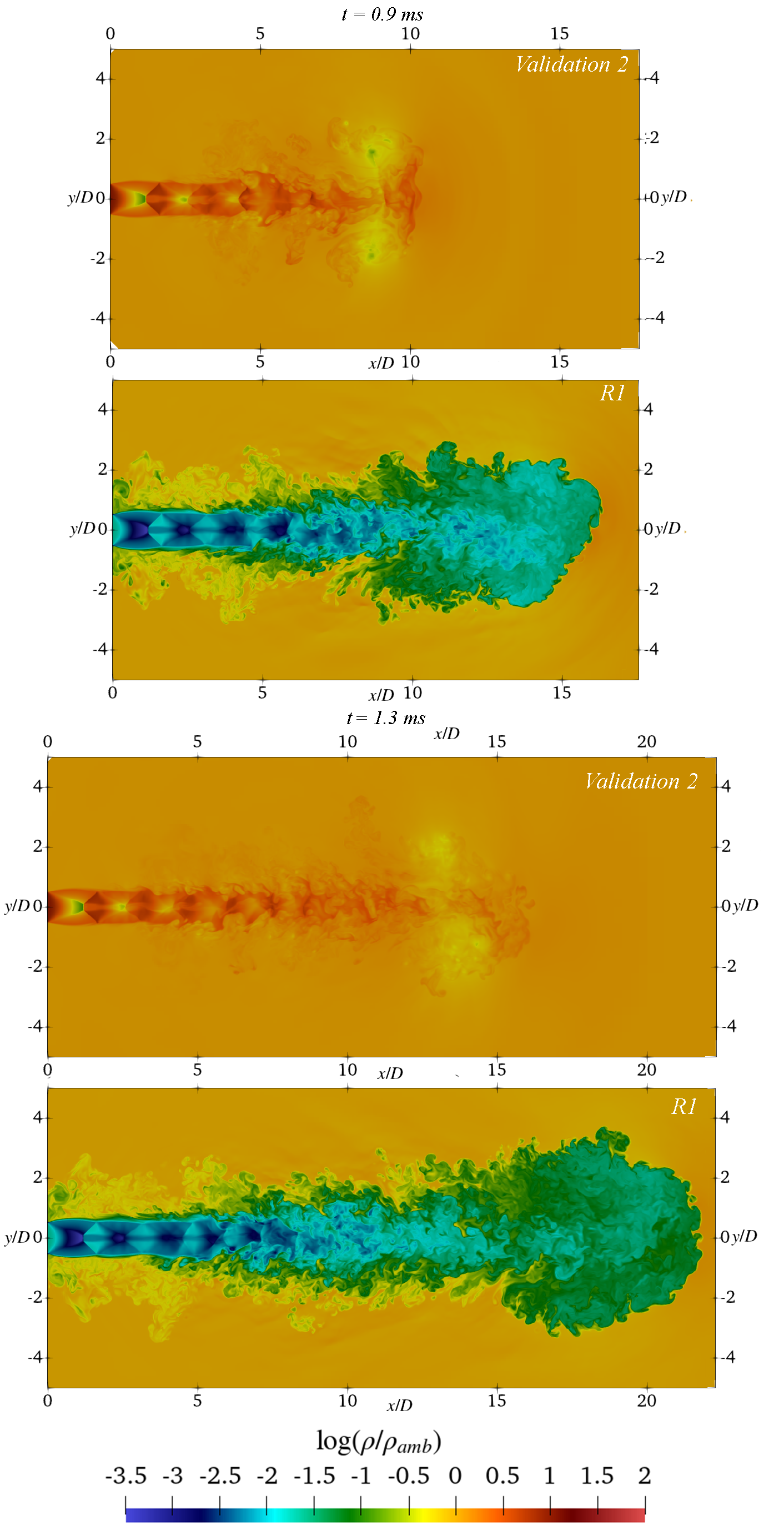}
\caption{Transient evolution of the jets of the validation case 2 and R1. Late stages.}
\label{fig:evCompAir2}
\end{figure}

With this section, we aim to investigate the differences in the flow structures of an under-expanded jet (and, consequently, mixing characteristics) depending on the chemical species injected.
Figures \ref{fig:evCompAir} and \ref{fig:evCompAir2} report the transient evolution of $\log(\rho/ \rho_{amb})$ for Validation 2 and R1 cases. The conditions for the two test cases are identical, except that hydrogen is injected instead of air for test R1. 

More precisely, in both cases, we observe the classical structures of under-expanded jets, such as barrel shock, cap flow discontinuity, oblique shock waves,  slipstream,  reflected shocks, and so on (see, e.g., \citet{2022-01-0505,zhang2019}).
It can be seen in the t = \SI{0.1}{\milli \second} snapshot of the Figure \ref{fig:evCompAir} a spherical propagating shock, the so-called bow shock. 
Vortex-induced shock pairs appear on the sides of the jet.

After \SI{0.3}{\milli\second}, we can observe a fully developed shock cell and the appearance of a normal shock. The triple point can be recognized at the intersection of the intercepting, normal, and reflected/oblique shocks. Slipstreams develop at this point: this is an embedded shear layer that divides the flow upstream of the Mach disc (subsonic) from the flow downstream of the reflected shock (supersonic).
Moving ahead in time, we can notice how this shock cell-based structure develops more and, at \SI{1.3}{\milli\second}, we can count two shock cells for the air jet while three shock cells for the H$_2$ jet.

\begin{figure}[h]
\centering
\includegraphics[width=\linewidth]{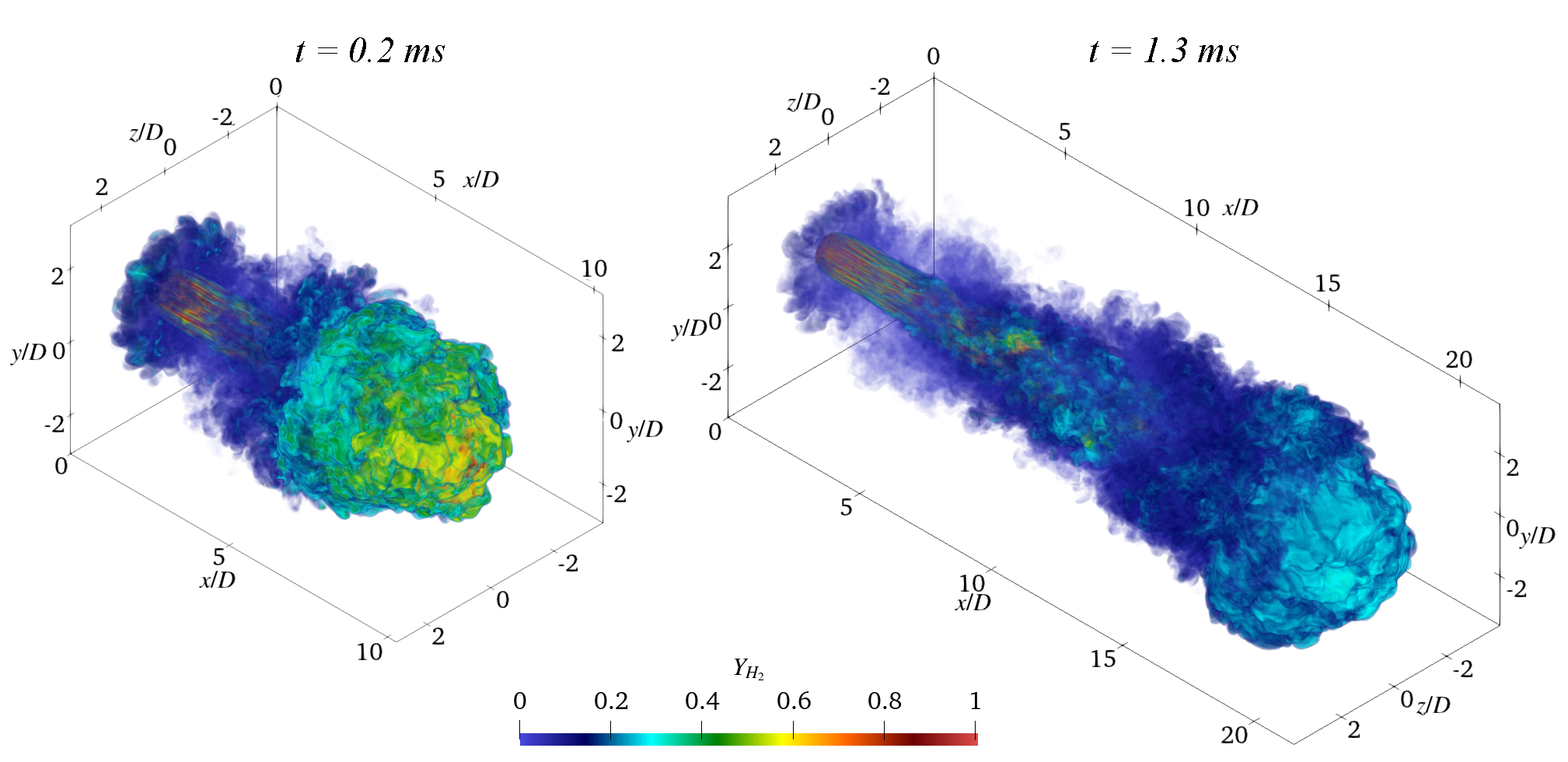}
\caption{Coanda effect for hydrogen jet.}
\label{fig:coanda}
\end{figure}

Nevertheless, we can underline several differences in the flow. First, the density of the H$_2$ jet is lower than the ambient one. This will have a strong influence on the development of the jet.
The initial vortex ring entraps much more surrounding fluid in the case of a hydrogen jet than with an air jet. A relevant amount of hydrogen moves aside on the exit section, producing remarkable vortical structures that create a hydrogen-enriched region in the radial direction (a sort of "Coanda effect"). Figure \ref{fig:coanda} highlights this phenomenon by showing the hydrogen mass fraction in the near nozzle zone for two different time steps. This phenomenon is observed only with hydrogen and not air; experimental and CFD studies confirm this behaviour \cite{Leick2023, MEROTTO2024284, 10.1007/978-3-658-45010-6_7}.

As the hydrogen jet evolves, we can observe the development and the evolution of a complex pattern of vortices (see figure \ref{fig:evCompAir} at $t=0.3 \, ms$ and $t=0.5\, ms$), which is absent with air jets, with the presence of a strong shear layer. This complex mixing layer is essential in chemically reactive flow since it promotes interaction with the environmental gas.

\begin{figure}[h!]
\centering
\includegraphics[width=\linewidth]{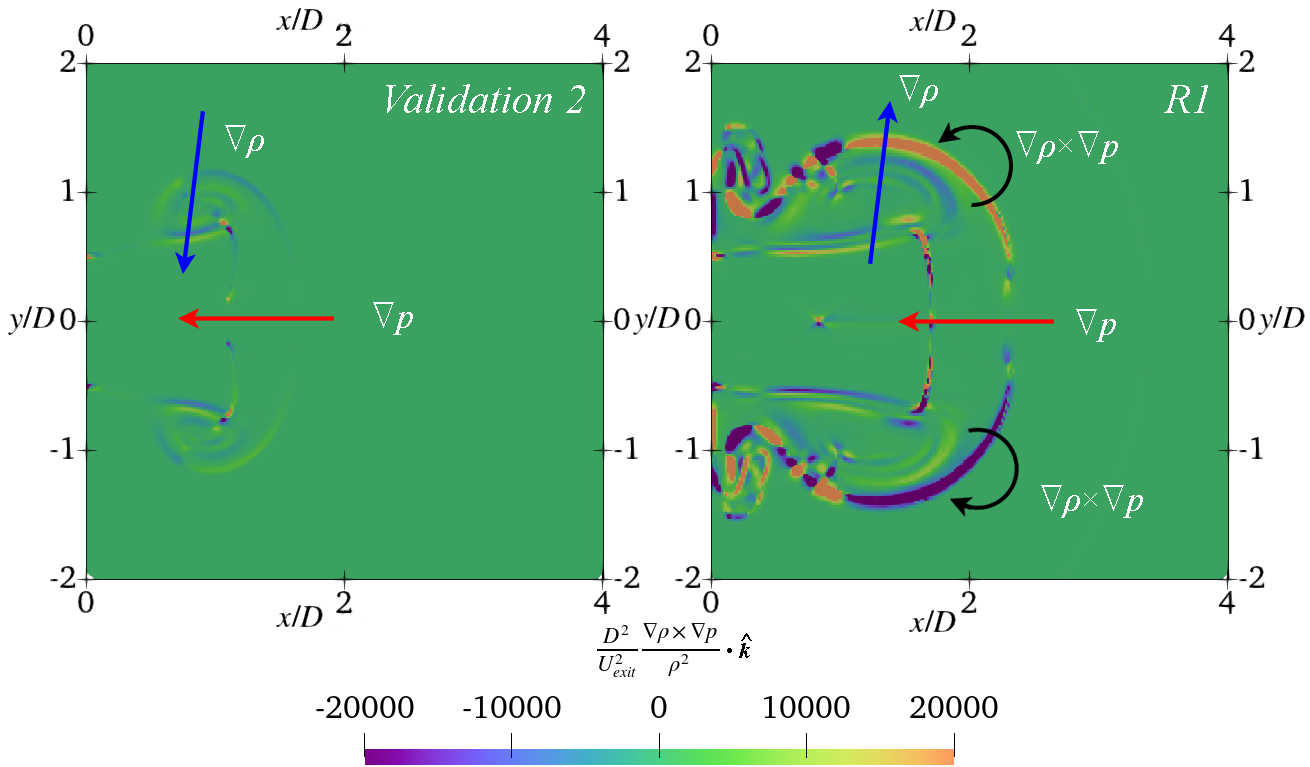}
\caption{Non-dimensional $z$ component of the baroclinic term in the vorticity equation on the plane $z=0$. Left: air-in-air jet; right: H$_2$-in-air jet}
\label{fig:baroclinicity}
\end{figure}

We can explain the differences between the two jets by examining the vorticity equation:
\begin{equation} 
\frac{\partial \vec{\omega}}{\partial t} + \vec{u} \cdot \nabla \vec{\omega}
= \vec{\omega} \cdot \nabla \vec{u} - \vec{\omega} ( \nabla \cdot \vec{u}) + \frac{1}{\rho^2} \nabla \rho \times \nabla p + \nu \nabla^2 \vec{\omega}
\label{vort}
\end{equation}
where we have the material derivative of the vorticity on the left-hand side of equation \ref{vort}, stretching and tilting in the first two terms on the right-hand side, the baroclinic term (the third term on the right), and the diffusion by viscosity in the last term.
The baroclinic term produces vorticity when the density gradient is not aligned with the pressure gradient; this term is always present in compressible flows, but we expect strong effects when injecting light gas (hydrogen, in the present simulations) into a heavy gas environment, like air-in-air or methane-in-air jets \cite{DURONIO2024109381}. Figure \ref{fig:baroclinicity} reports the $z$ component of this term in non-dimensional form for the validation case 2 and R1 jets.

\begin{figure}[h]
\centering
\includegraphics[width=.95\linewidth]{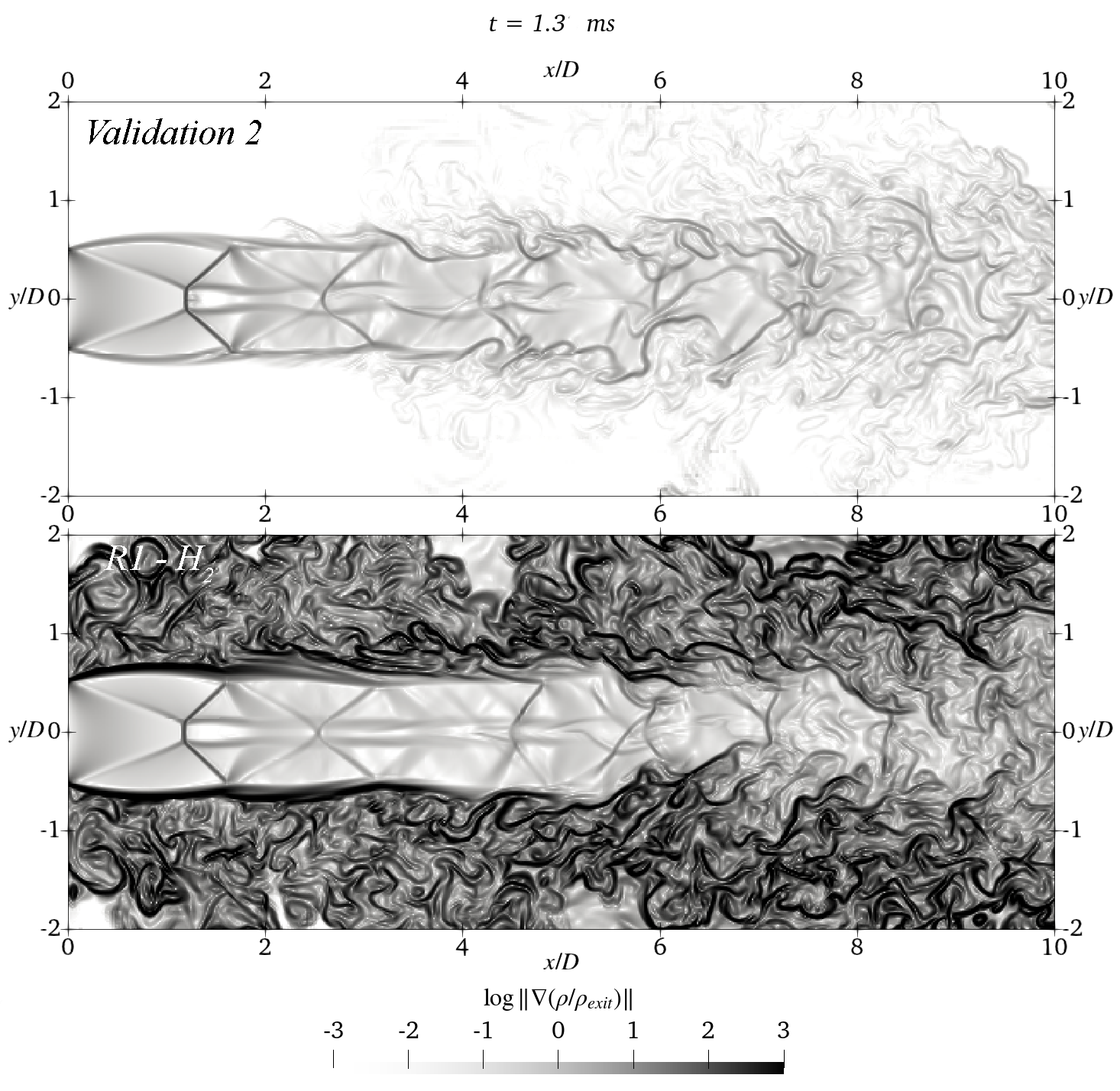}
\caption{ Validation case 2 (top) and R1 (bottom). The logarithm of the non-dimensional density gradient.}
\label{fig:machcomp}
\end{figure}

For the H$_2$-in-air jet, the pressure gradient $\nabla p$ is almost parallel to the $x$ axis and directed to the left, while the density gradient is pointing outward because the jet core is occupied by (the lighter) hydrogen, as shown by the arrows on the right of figure \ref{fig:baroclinicity}; therefore, the baroclinic term in the vorticity equation induces counterclockwise rotation in the upper part of the figure and clockwise rotation in the lower portion (as shown by the circular arrows in the picture) that push the hydrogen outward. 
On the contrary, in the air-in-air flow, the density gradient points toward the axis; consequently, the induced rotation is opposite to the one observed with the hydrogen flow. In addition, the magnitude of this term is much lower. 
This baroclinic term justifies the completely different morphology of the outer jet between the single-species and the two-species flow. Nonetheless, the shock structure in the core is remarkably similar. Figure \ref{fig:machcomp} is a zoomed view of the Mach disc for the two investigated cases performed after \SI{1.3}{\milli\second}.

\begin{figure}[h!]
\includegraphics[width=.62\linewidth]{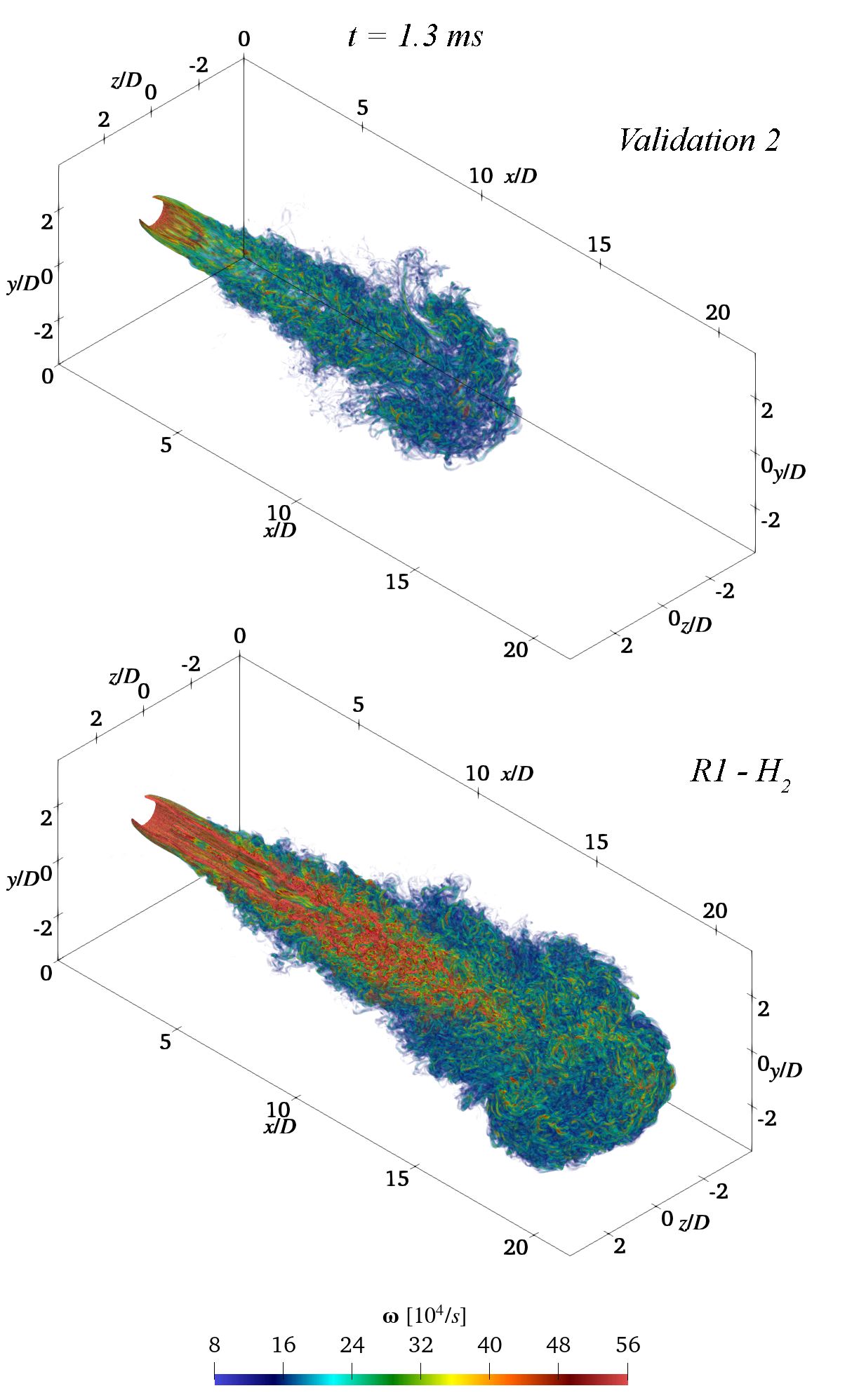}
\caption{ Magnitude of vorticity vector for the air and hydrogen jet with pressure ratio = 4.2.}
\label{fig:vort}
\end{figure}

The figure shows that the structure of the primary shock cell is not affected by the species injected, as both the converging shocks and the normal shock have a similar morphology. As previously reported, the two-species jet exhibits a broader shear layer outside the jet core, where the hydrogen quickly mixes with the surrounding air. To quantify this aspect, we reported in Figure \ref{fig:vort}  the magnitude of the vorticity vector after \SI{1.3}{\milli\second}. 

\begin{figure}[h!]
\centering
\includegraphics[width=.62\linewidth]{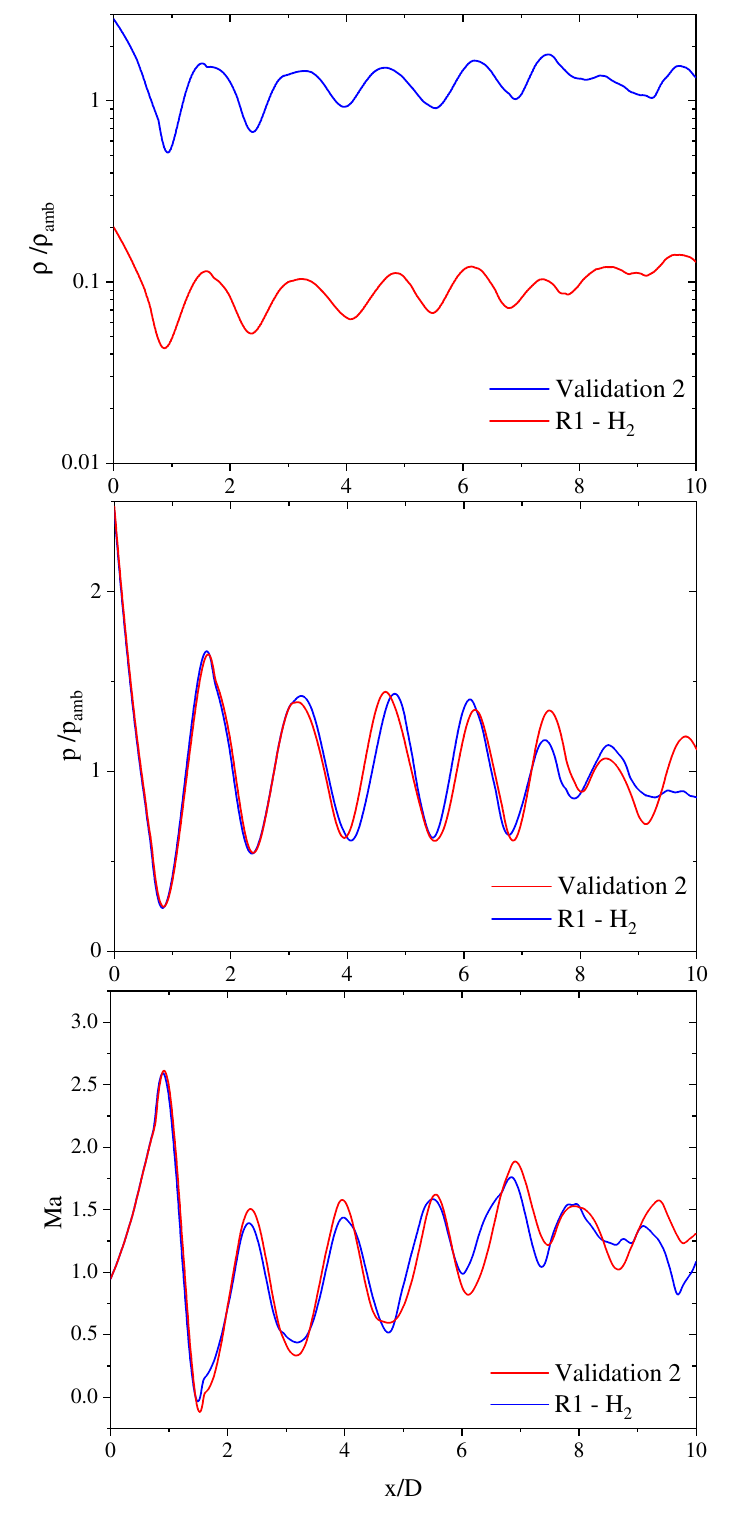}
\caption{Density, pressure, Mach number axial plots for air and H${_2}$ jets with PR=4.2. }
\label{fig:axial}
\end{figure}

The plot is a volumetric rendering where we removed a quarter of the jet to show its internal structure. 
The vorticity is higher for the hydrogen jet. In the near nozzle zone, both the jets present streamwise vortices outside the potential core in the shear layer with small-scale vortices downstream. 
As already underlined, the vorticity field provides an overview of the turbulence which is relevant in chemically reactive flows since it drives the mixing of the air with the fuel and so the efficiency of the subsequent combustion process \cite{BUTTAY20188488,Hamzehloo2016b}.
A more detailed quantitative comparison of the two jets is reported in the figure \ref{fig:axial}, where the axial plots of non-dimensional density, pressure, and Mach number fields are reported.

As already discussed, relevant differences can be recognized in the density field.
\subsection{Effects of the pressure ratio on the H${_2}$ jets}
\begin{figure}[h!]
\centering
\includegraphics[width=\linewidth]{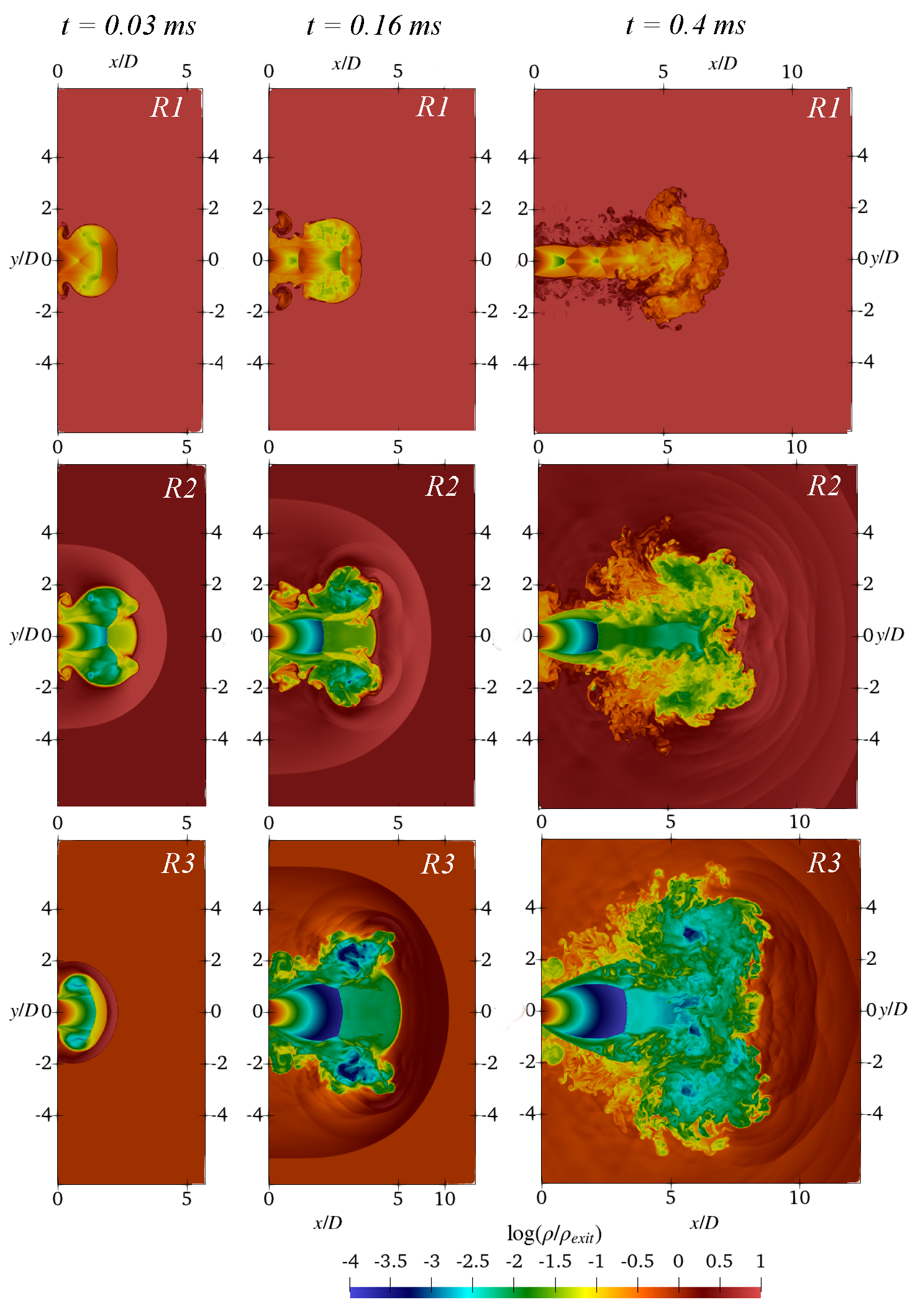}
\caption{Transient evolution of the hydrogen jets R1, R2, and R3.}
\label{fig:transPR}
\end{figure}

Figure \ref{fig:transPR} reports the early stages of the jets R1, R2, and R3 evolution, where the pressure ratio equals 4.2, 12.6, and 25.2, respectively.

\begin{figure}[h!]
\centering
\includegraphics[width=.8\linewidth]{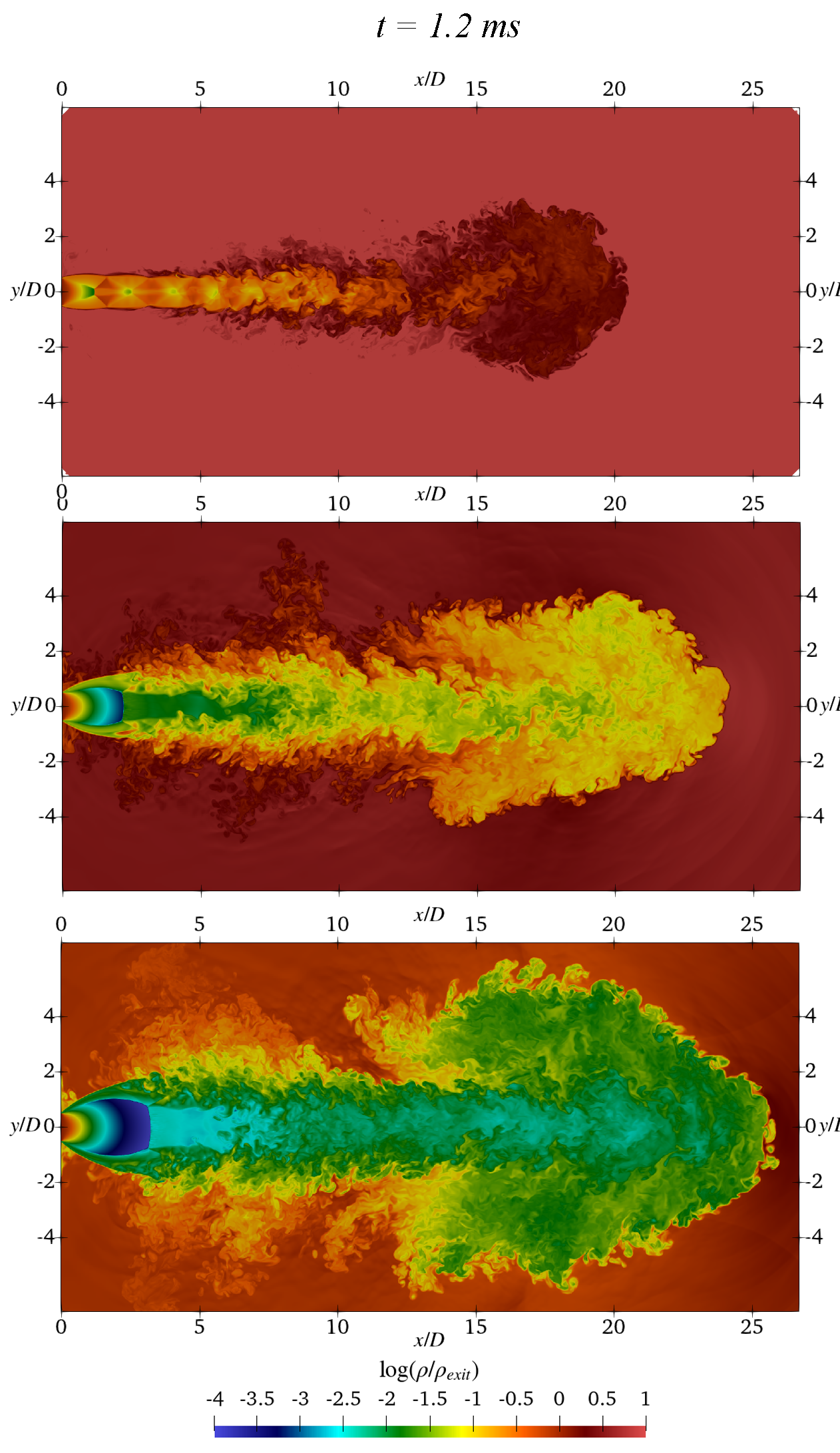}
\caption{Transient evolution of the hydrogen jets R1, R2, and R3, $t=\SI{1.2}{\milli\second}$.}
\label{fig:transPR2}
\end{figure}

The early stages of the jets significantly differ when increasing the pressure ratio. The jet with the lower pressure ratio shows two converging shocks that merge at a point on the jet's axis and a pronounced primary vortex ring. When increasing the upstream pressure, the classical Mach disc shock appears, and the structure of the vortex rings changes. We can observe a primary vortex ring and a significant secondary vortex ring.  A cap flow discontinuity delimits a distinguishable zone downstream of the Mach disc shock.
The pressure ratio also has a significant effect on the shear layer. Indeed, looking at the snapshots in Figure \ref{fig:transPR}, it can be concluded that a higher injection pressure promotes a broader shear layer, enhances the Coanda effect, and, as discussed below, helps form an ignitable air/fuel mixture.
At t = \SI{1.2}{\milli\second} (figure \ref{fig:transPR2}), the final under-expanded structures that characterized the jet can be observed.

Following the classification proposed by \citet{franquet2015free} and Duronio \textit{et al.} \cite{en16186471}, the jet of case R1 is highly under-expanded, while the jets R2 and R3 are extremely (or very highly) under-expanded. Indeed, in the first case, the jet has a “barrel” structure that repeats three times downstream. The triple point can be recognised as the intersection of the intercepting shock, the Mach disc, and the reflected shock.
When increasing the injection pressure, the structure of the jet is dominated by a single barrel shock. A normal shock no longer characterises the Mach disc, but a relevant curvature appears.
The Mach disc's width and height measurements are reported in Table \ref{tab:machDh} for the three pressure ratios investigated.  
\begin{table}[h]
\caption{\label{tab:machDh}Mach disc's width and height measurements. Cases R1, R2 and R3. }
\begin{ruledtabular}
\begin{tabular}{ccccc}
Jet \#	& R1  & R2 & R3 \\ \hline
Height [-] & 1.17D & 2.25D & 3.3D\\ 
Width [-] & 0.2D & 1.06D & 1.5D\\ 
\end{tabular}
\end{ruledtabular}
\end{table}

\begin{figure}[h!]
\centering
\includegraphics[width=.8\linewidth]{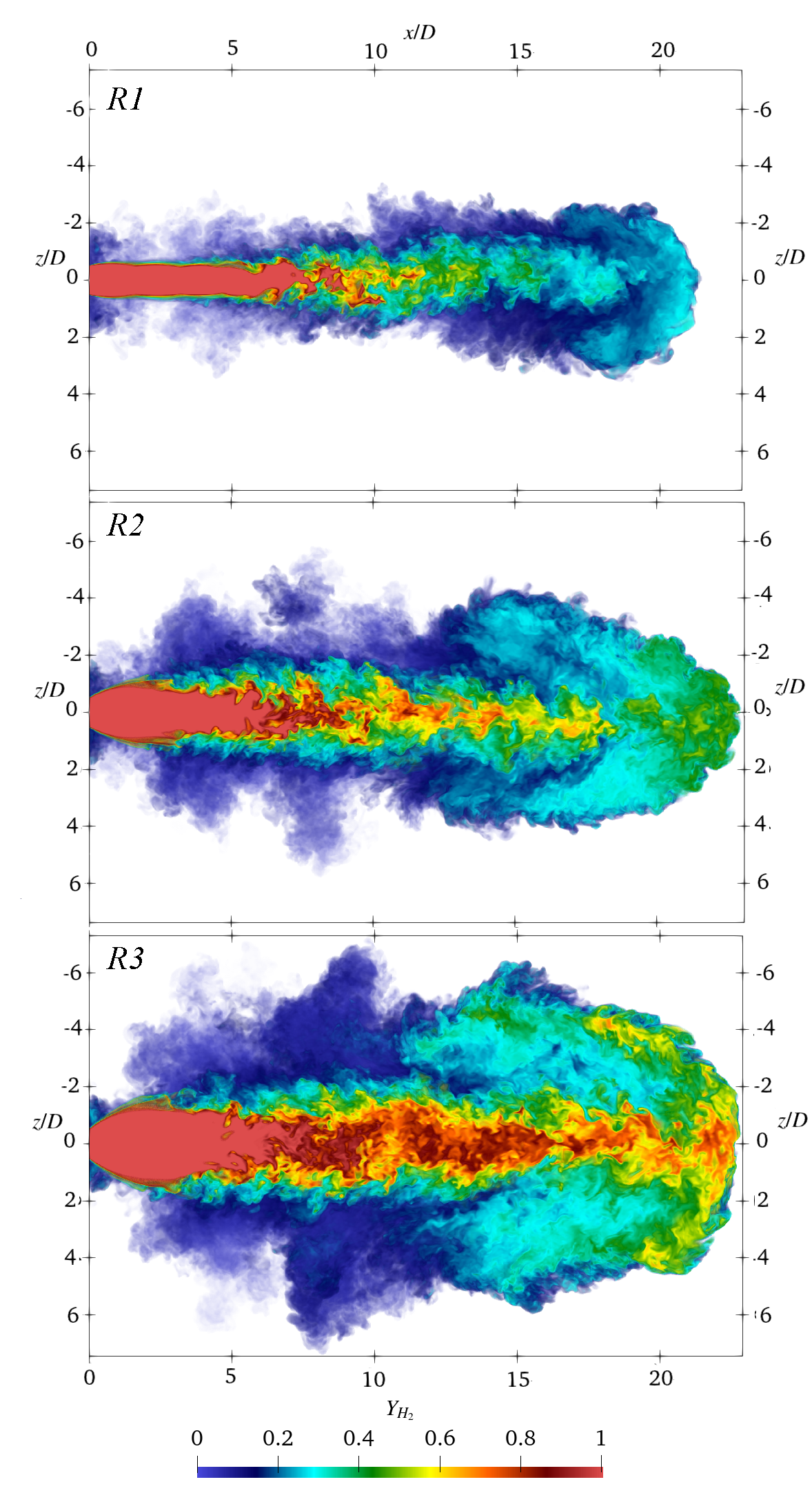}
\caption{Hydrogen mass fraction for jets R1, R2, and R3, $t=\SI{1.1}{\milli\second}$.}
\label{fig:yh2axis}
\end{figure}

The lack of a normal shock sequence promotes jet penetration and reduces the so-called potential core's length, increasing the mixing between hydrogen and air; this aspect is shown in figure \ref{fig:yh2axis}, where it can be observed that the zone of the jet characterized by the sequence of barrel shocks does not show any mixing activity, which is completely inhibited. The increase in the pressure ratio reduces the length of this zone.

\begin{figure}[h!]
\includegraphics[width=\linewidth]{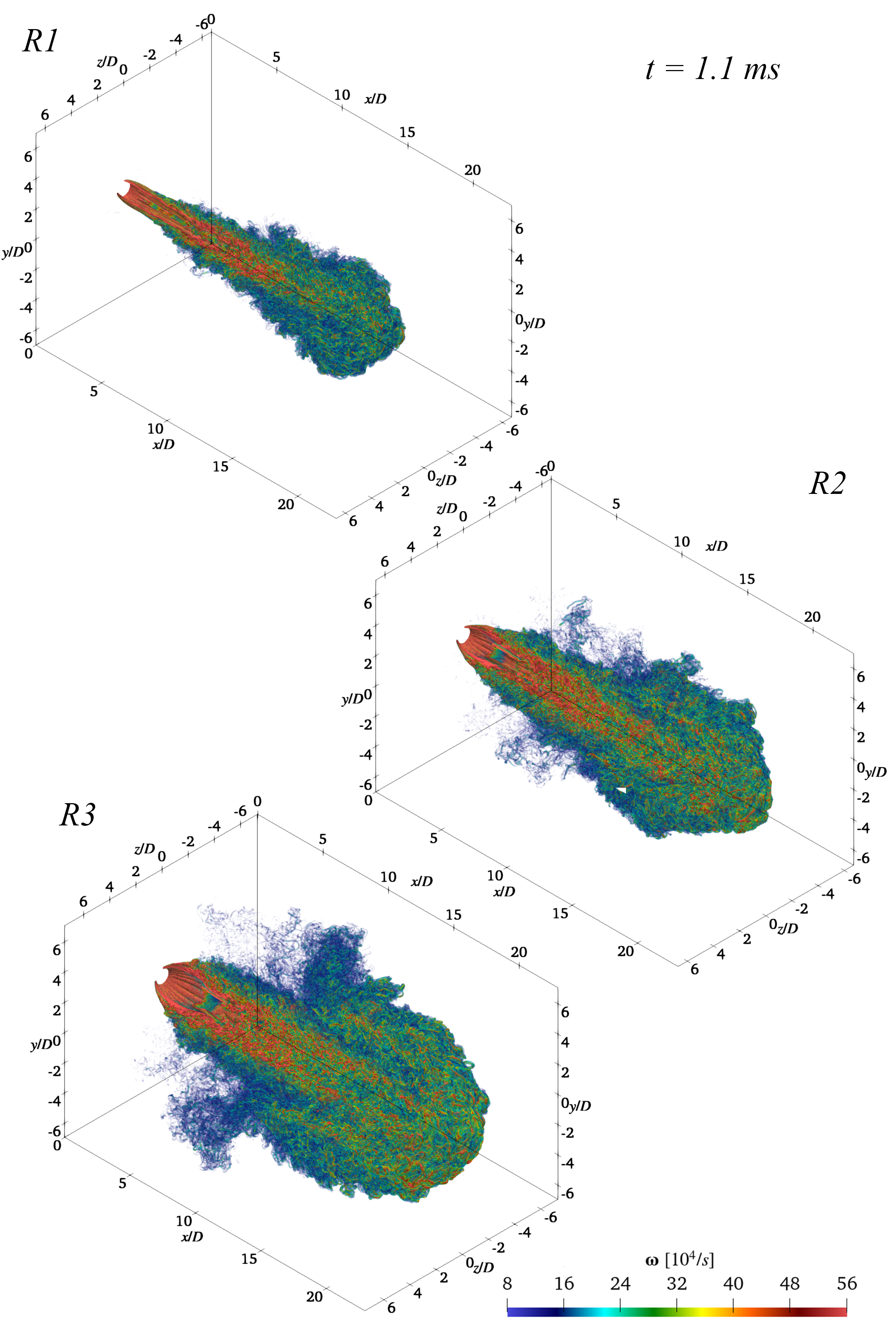}
\caption{Magnitude of the vorticity vector for cases R1, R2, and R3}
\label{fig:vortH2}
\end{figure}

Figure \ref{fig:vortH2} shows a volumetric rendering of the vorticity magnitude for the three cases.

Vorticity is an important parameter in assessing the mixing activity: the higher the pressure ratio, the stronger the vorticity.

\begin{figure}[h!]
	\begin{subfigure}[b]{0.35\textwidth}
\includegraphics[width=\linewidth]{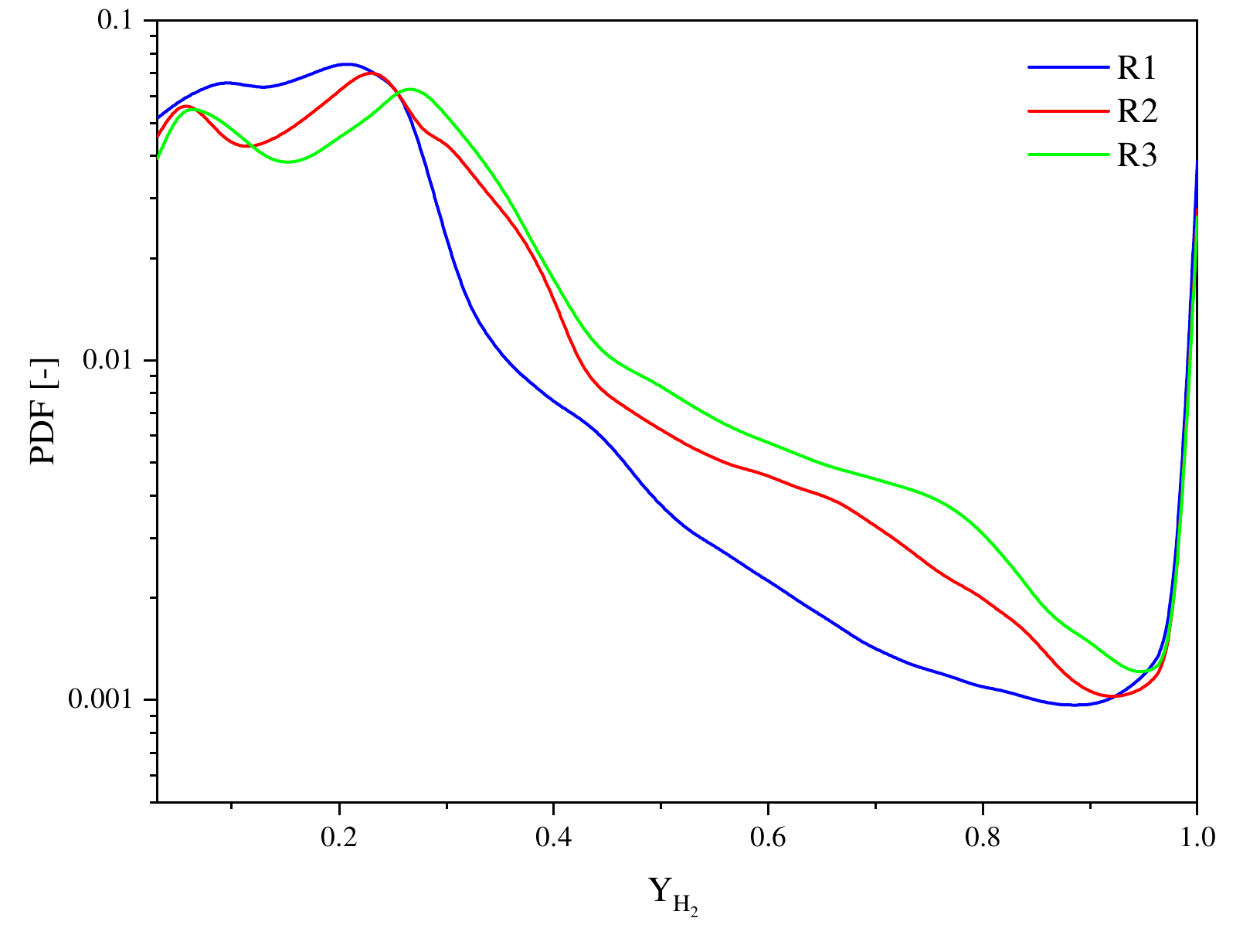}
	\end{subfigure}
	\begin{subfigure}[b]{0.35\textwidth}
\includegraphics[width=\linewidth]{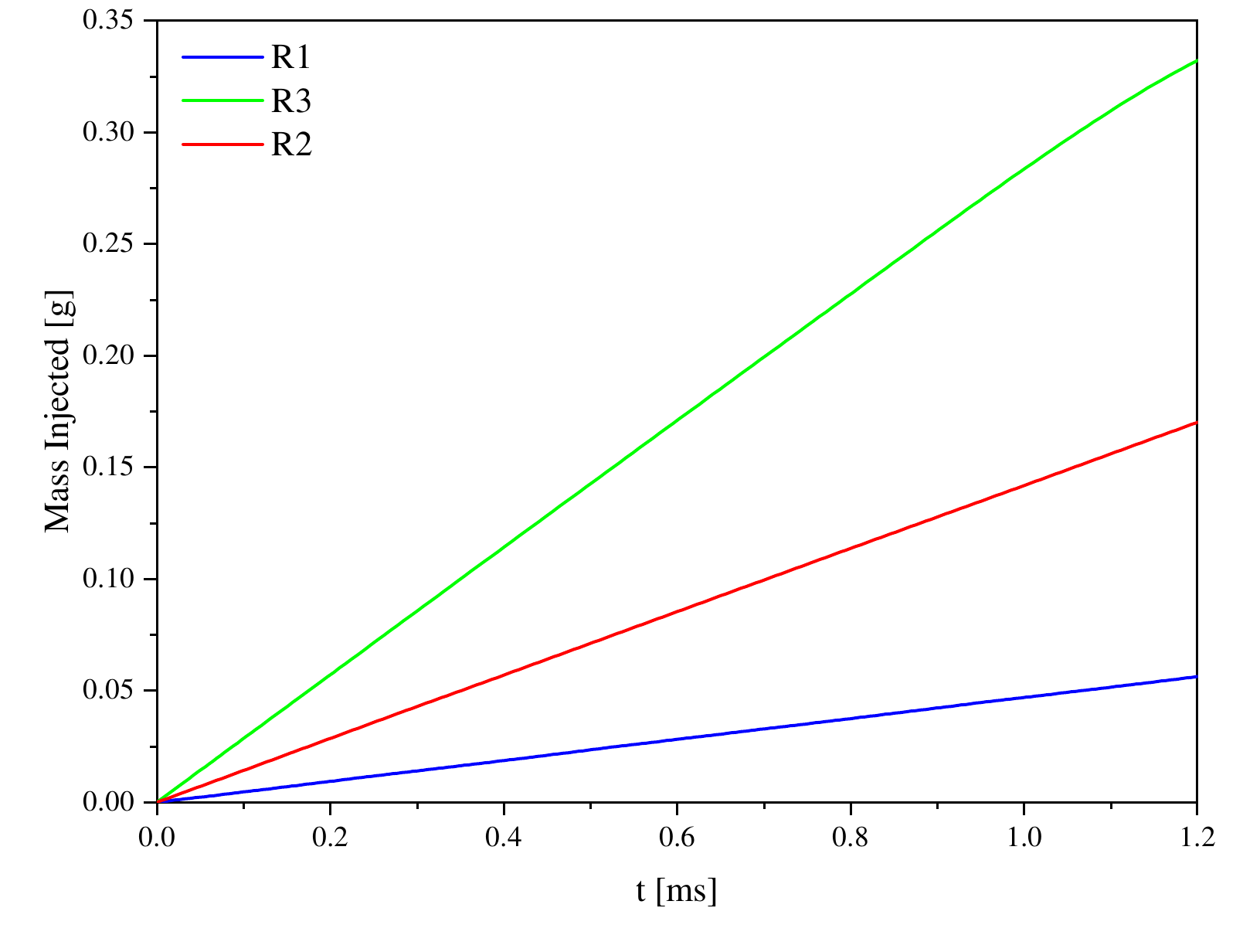}
	\end{subfigure}\caption{Probability distribution of $H_2$ mass fraction at \SI{1.2}{\milli\second} and hydrogen mass injected in dimensional terms.}
\label{fig:pdfpr}
\end{figure}

Obtaining a quantitative evaluation of the overall quality of the mixture is essential when dealing with combustion processes. Thus, we defined a discrete mass-weighted probability density function as:
\begin{equation}
    \begin{array}{l}
	PDF_k(Y_{H_2})= \\*[3mm]
    \displaystyle
    \frac{1}{M_{tot}} \sum_{i=1}^{N} \left\{
	\begin{array}{ll}
        \displaystyle
		\rho_i \, \delta V_i \, {Y_{H_2}}_{,i} 
        & \mbox{for}  \quad k/K \leq {Y_{H_2}}_{,i} \leq (k+1)/K \\
		0 &  \mbox{otherwise} 
	\end{array} \right.
    \end{array}
	\label{pdf}
\end{equation}
where:
\begin{itemize}
 \item[-] $PDF_k(\cdot)$ is the probability density function;
    \item[-] N is the total cell count;
    \item[-] $K$ is the number of partitions; in this case K=100;
    \item[-] $0 \le k < 100$;   
    \item[-] $\delta V_i$ is the cell volume;
    \item[-] $\rho_i$ is the mixture density;
    \item[-] ${Y_{H_2}}_{,i}$ is the hydrogen mass fraction in the $i$-th cell;
    \item[-] $M_{tot}$ is the total injected mass at a certain instant;
\end{itemize}
Figure \ref{fig:pdfpr} reports the probability density function distribution and the mass injected for the three cases investigated at \SI{1.2}{\milli\second}.

The PDFs for the three cases are very similar, showing a peak around $Y_{H_2}=0.2$; the amount of hydrogen mass injected grows with the injection pressure.

\subsection{Effects of the nozzle shape}
\begin{figure}[h!]
\centering
\includegraphics[width=\linewidth]{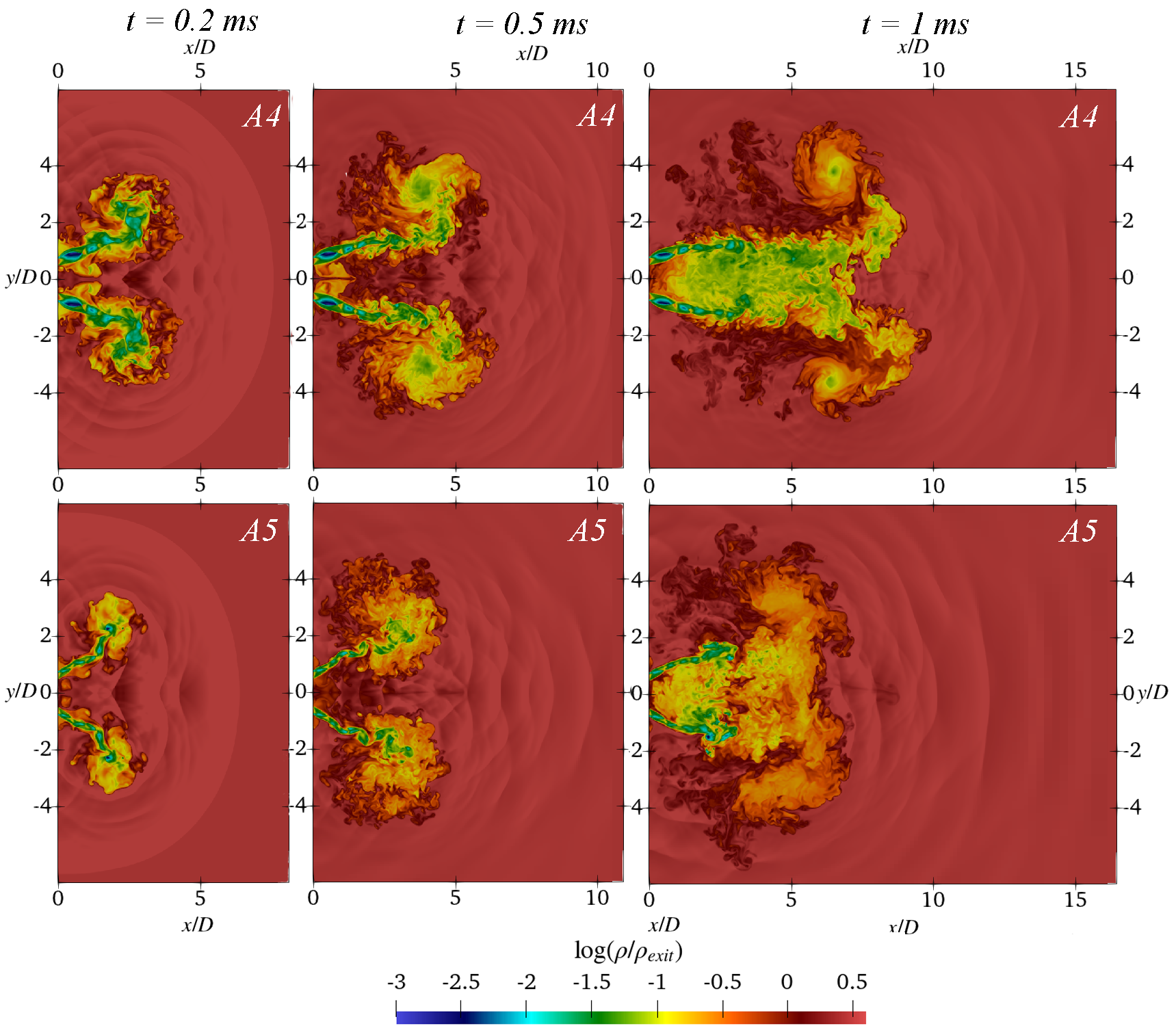}
\caption{ Early stages of the transient evolution for the jets A4 and A5. Plot of the logarithm of the non-dimensional density gradient. }
\label{fig:compAnn1}
\end{figure}

The nozzle design for gaseous injectors is essential to developing efficient injection devices. For this reason, we investigate two annular nozzles characterised by jet angles equal to 90$^\circ$ (case A4) and 135$^\circ$ (case A5).
Figure \ref{fig:compAnn1} compares the jet's transient evolution for the early stages of the simulated transient. As before, we plotted the $\log(\rho/\rho_{exit})$.

\begin{figure}[h!]
\centering
\includegraphics[width=\linewidth]{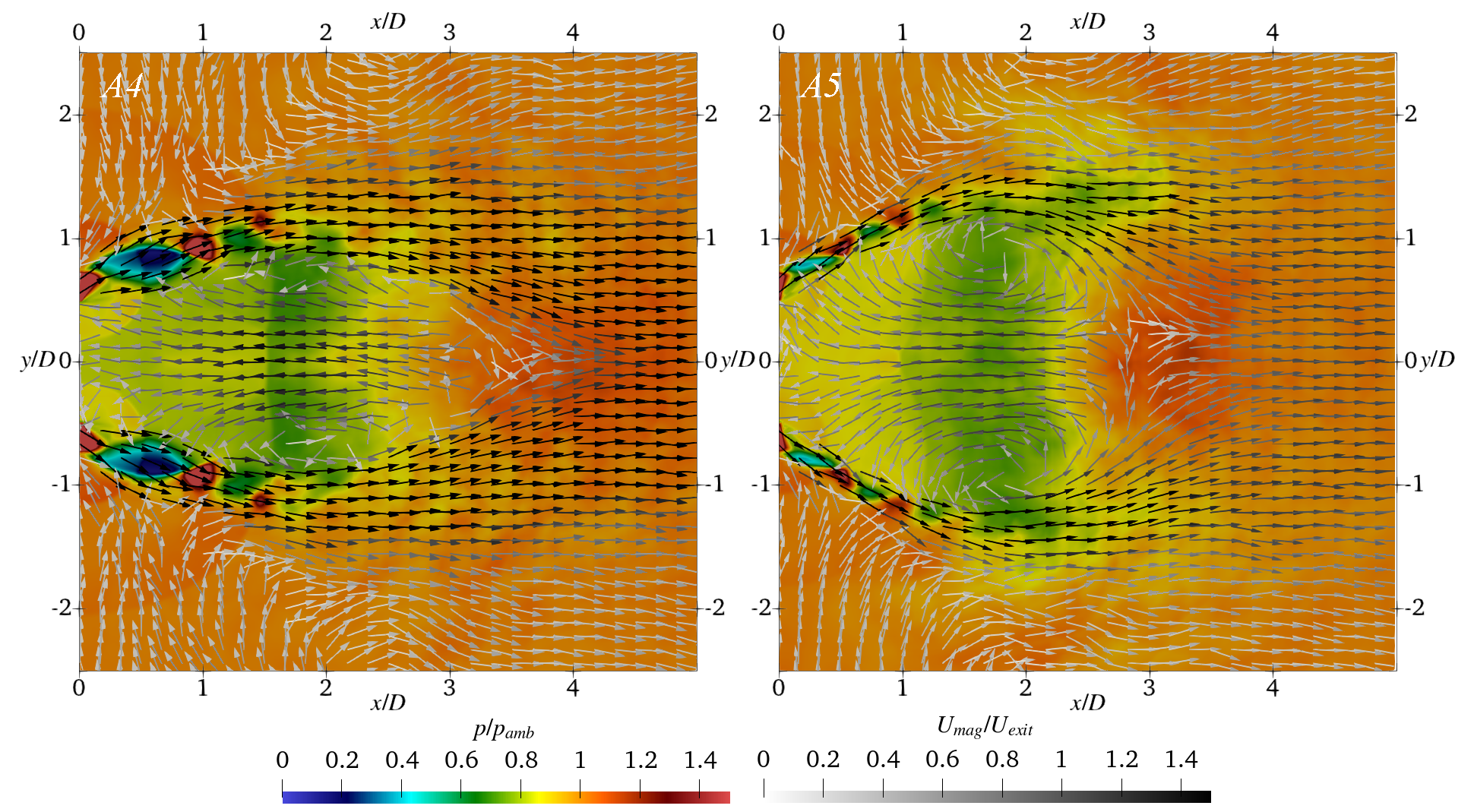}
\caption{Time averaged pressure field and velocity vectors.}
\label{fig:pressVel}
\end{figure}

The evolution is very different when compared with the round jet.  At t =\SI{0.2}{}$\div$\SI{0.5}{\milli\second}, the hollow-cone jets present a conical structure characterized by the angle imposed. The pressure inside the cone becomes gradually lower than the external pressure. The jet continues its evolution by contracting towards the x-axis, as shown in figure \ref{fig:pressVel}, where we reported the time-averaged velocity vectors and the pressure field over $0\leq t \leq \SI{2.4}{\milli\second}$ on the axial plane.

\begin{figure}[h!]
\includegraphics[width=.8\linewidth]{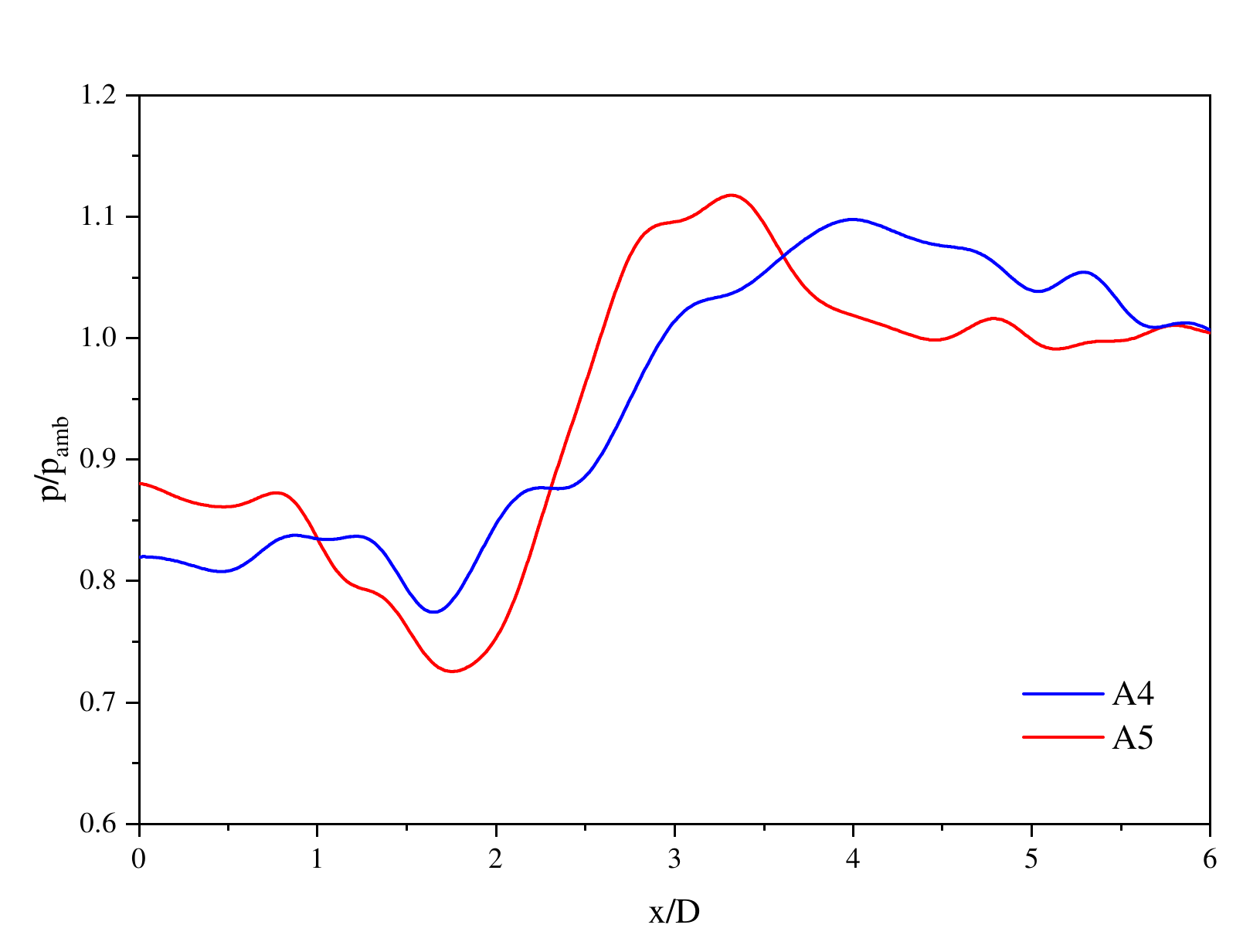}
\caption{Pressure profiles along the jet's axis for cases A4 and A5.}
\label{fig:axialpressure}
\end{figure}

Outside the core, in the near nozzle zone, air cannot enter within the jet due to a series of shock cells. On the other hand,  we can observe a series of vortices rotating toward the central axis that drag the hydrogen into the jet's core.  This flow path extends for approximately 4D downstream of the nozzle exit section. The intensity of this phenomenon increases with the jet cone angle.
Figure \ref{fig:axialpressure} reports a quantitative evaluation of the pressure diminishing.

\begin{figure}[h!]
\includegraphics[width=\linewidth]{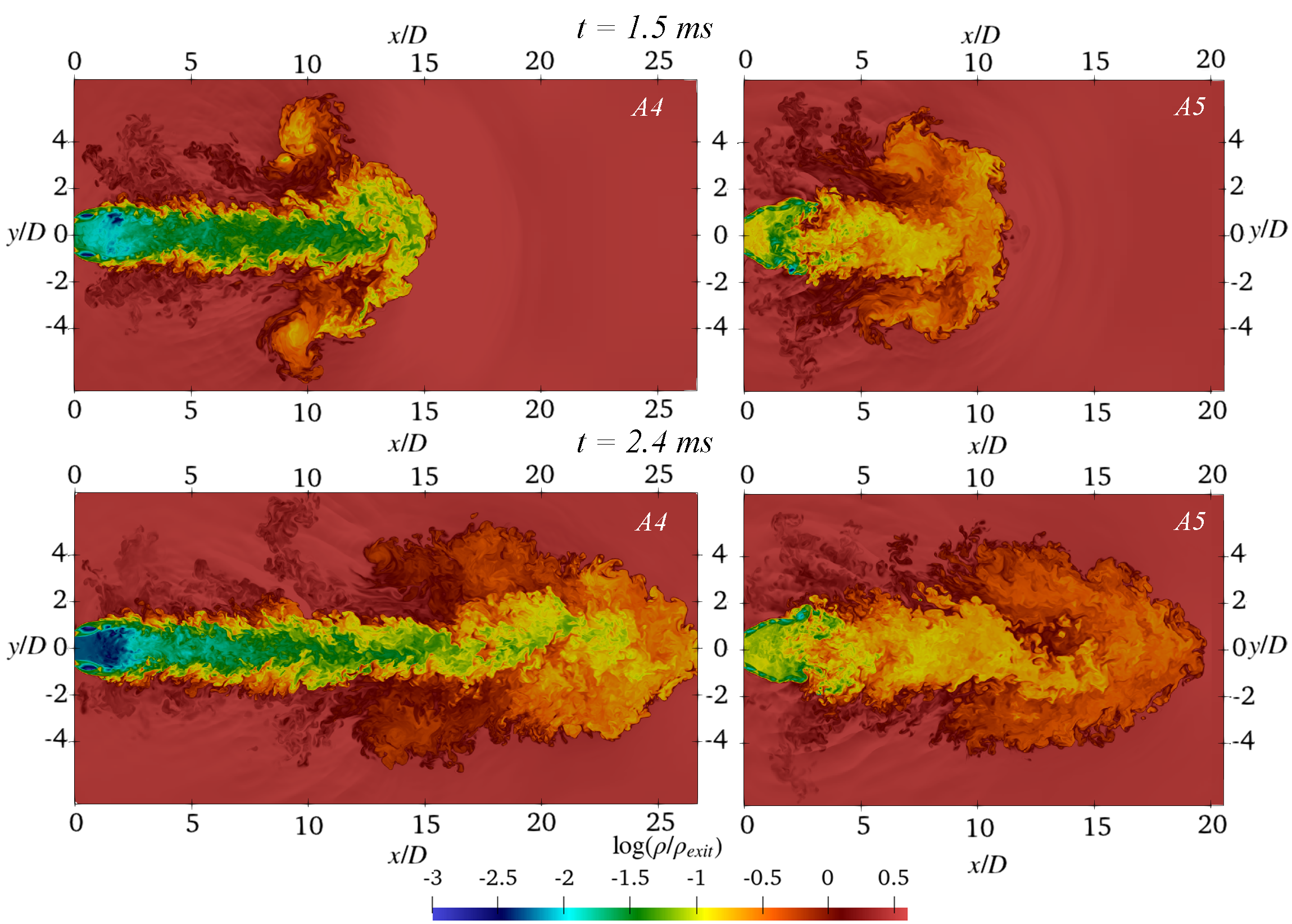}
\caption{Transient evolution of the jets A4 and A5. Late stages.}
\label{fig:compAnn2}
\end{figure}

The pressure reduction reaches 30\% in the case A5, while in the case A4 it is at most 20\%.
As illustrated by Figure \ref{fig:compAnn2}, these vortical structures remain active until the jet spreads in the surroundings.

The subsequent injection stages are characterised by the jet advancing toward the right boundary of the domain with a velocity greater for jet A4 because of the smaller cone angle. Although the shape of the jet is similar, the cone angle modifies the hydrogen concentration, as shown in the figure by $\log(\rho/\rho_{exit})$.
Various experimental investigations confirm this evolution mechanism, and it is characteristic of hollow cone jets \cite{LEE20214538,CORATELLA2024432,2024-01-2617}.

\begin{figure}[h!]
\includegraphics[width=\linewidth]{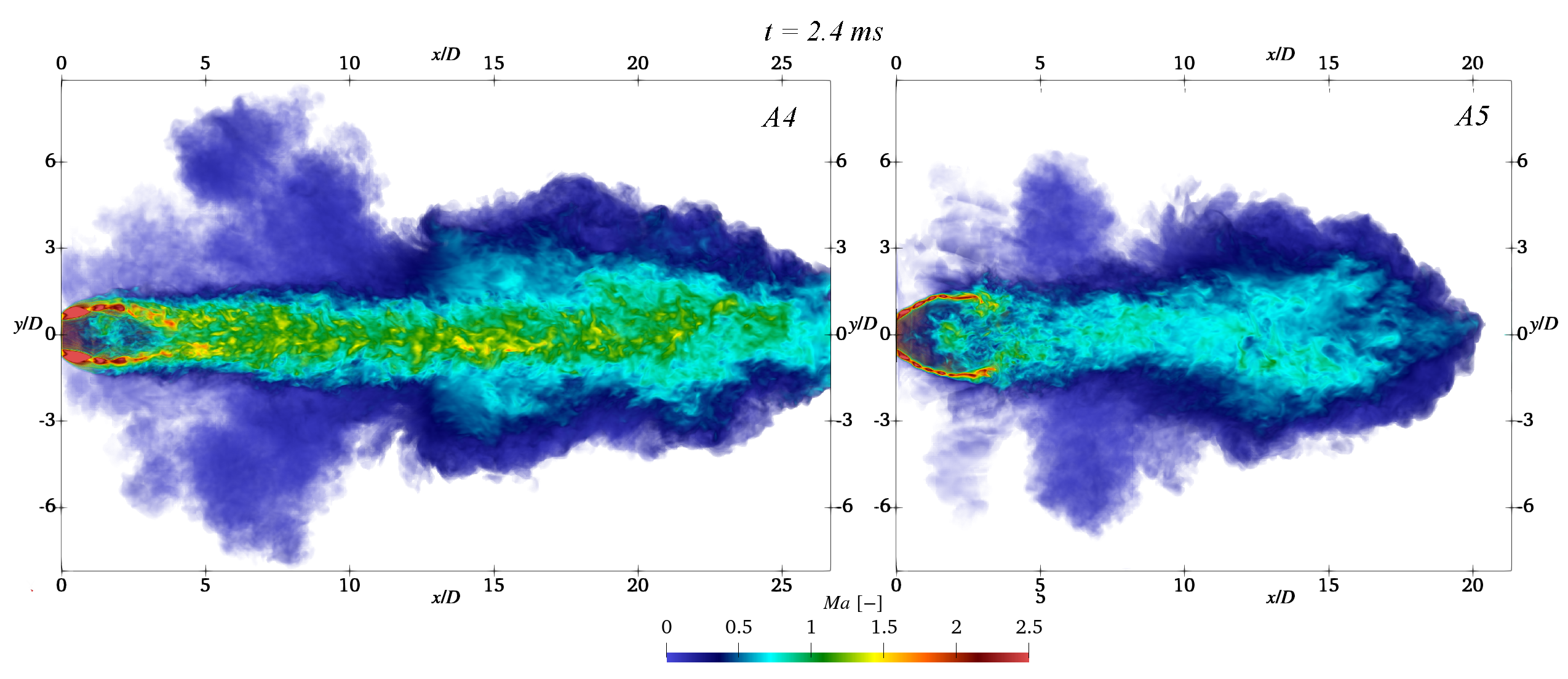}
\caption{Mach number plot for cases A4 and A5, t = \SI{2.4}{\milli\second}.}
\label{fig:mach}
\end{figure}

Figure \ref{fig:mach} shows the Mach number for the two jets at t = \SI{2.4}{\milli\second}.

The hollow conical structure of the jet determines a ring-shaped spatial arrangement of the shock waves. The single Mach disc observed with the circular nozzle disappears, but barrel shocks also appear in test cases in A4 and A5; moreover, their size diminishes with the increasing jet angle. The number of these shock cells is larger for case A5. We can conclude that, when performing a hollow cone injection, the increase of the jet angle reduces the Mach number and the hydrogen velocity; therefore,  with the same nozzle area, the injection of the same amount of $H_2$ requires a longer time. 

\begin{figure}[h!]
\centering
\includegraphics[width=.7\linewidth]{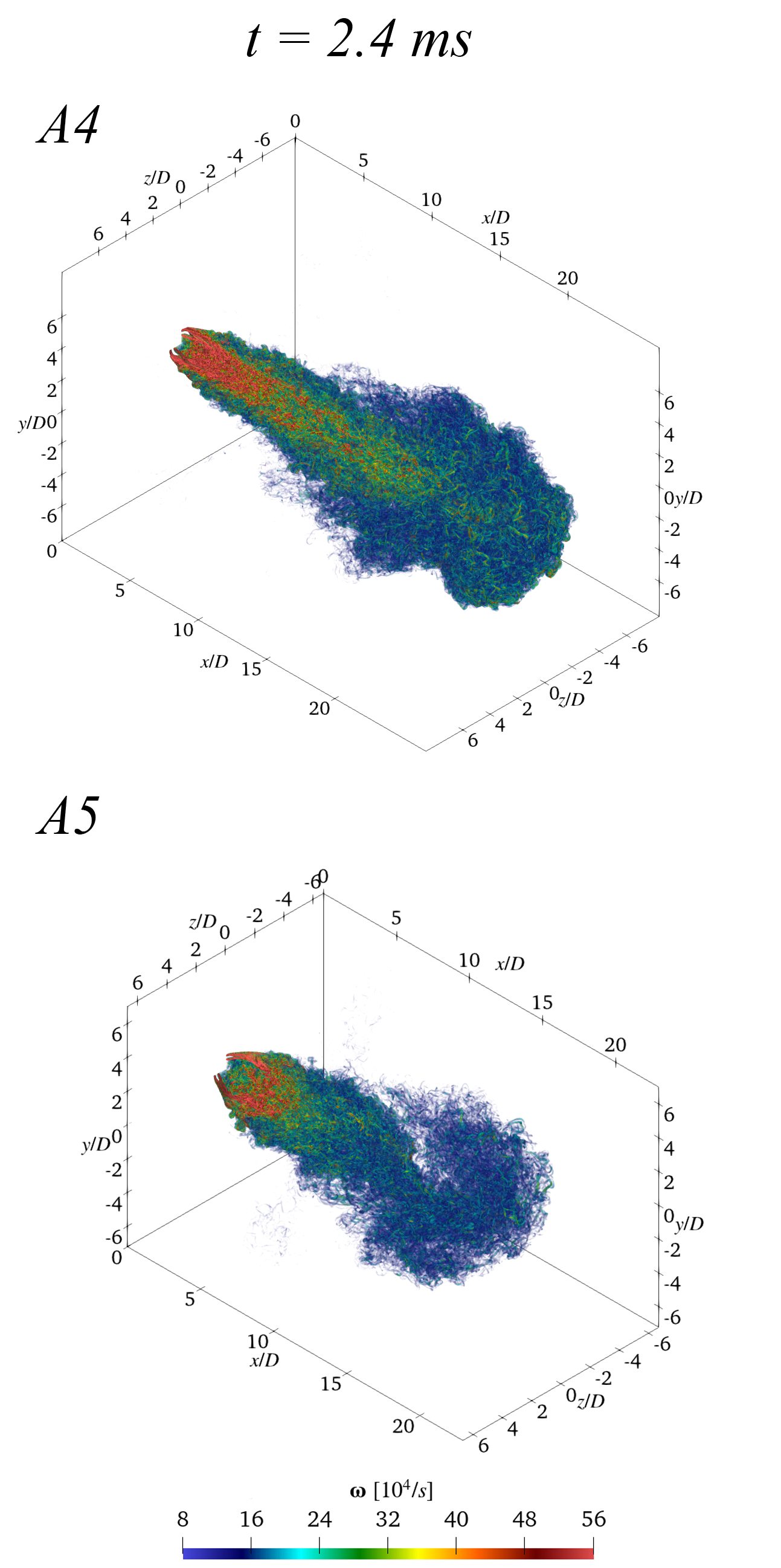}
\caption{Magnitude of the vorticity vector for cases A4 and A5, t = \SI{2.4}{\milli\second}.}
\label{fig:vort2}
\end{figure}

Figure \ref{fig:vort2}  shows a volumetric rendering of the vorticity magnitude for the cases A4 and A5 at t = \SI{2.4}{\milli\second}.

\begin{figure}[h!]
	\begin{subfigure}[b]{0.35\textwidth}
\includegraphics[width=\linewidth]{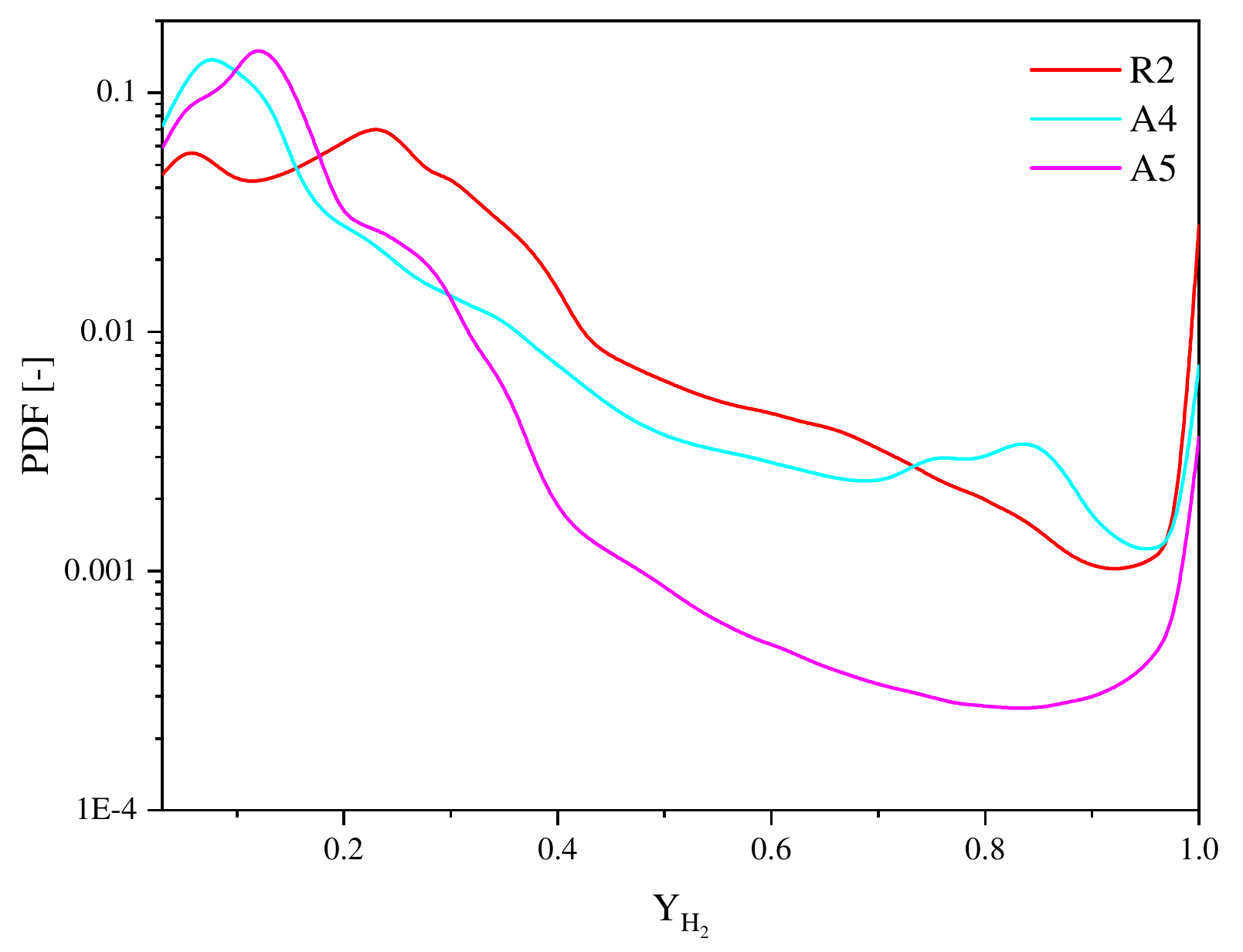}
	\end{subfigure}
	\begin{subfigure}[b]{0.35\textwidth}
\includegraphics[width=\linewidth]{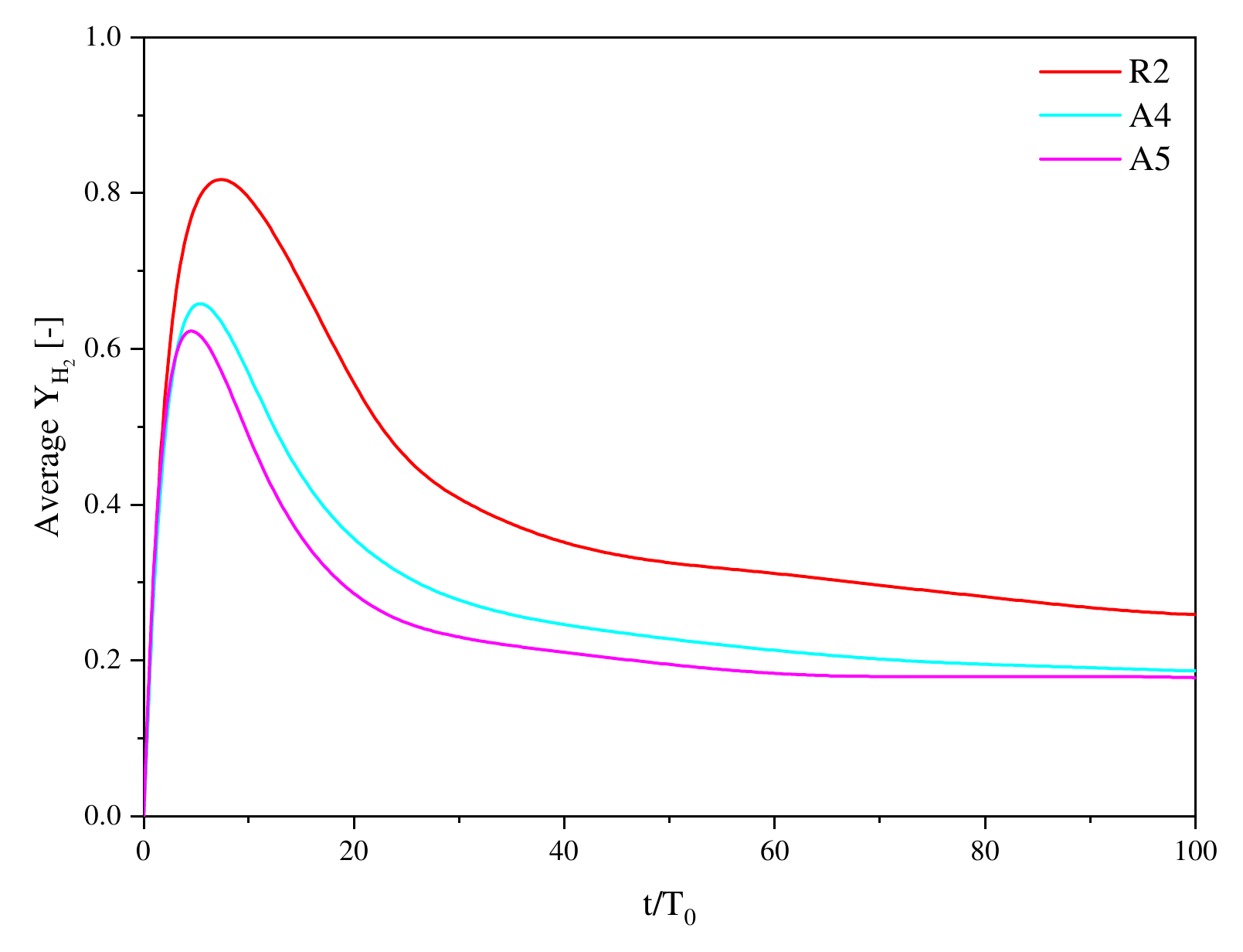}
	\end{subfigure}\caption{Probability distribution of $H_2$ mass fraction and average $H_2$ mass fraction as a function of time.}
\label{fig:pdfAnnular}
\end{figure}

To quantitatively assess the mixture quality obtainable with a hollow cone injection, in Figure \ref{fig:pdfAnnular} we plotted the PDF and the spatial average $H_2$ concentration over the domain as a function of time for the cases R2, A4, and A5. To have comparable results, we plotted the PDF at the time instant required to have the same injected mass equal for all the cases.

\begin{figure*}[t]
\includegraphics[width=\linewidth]{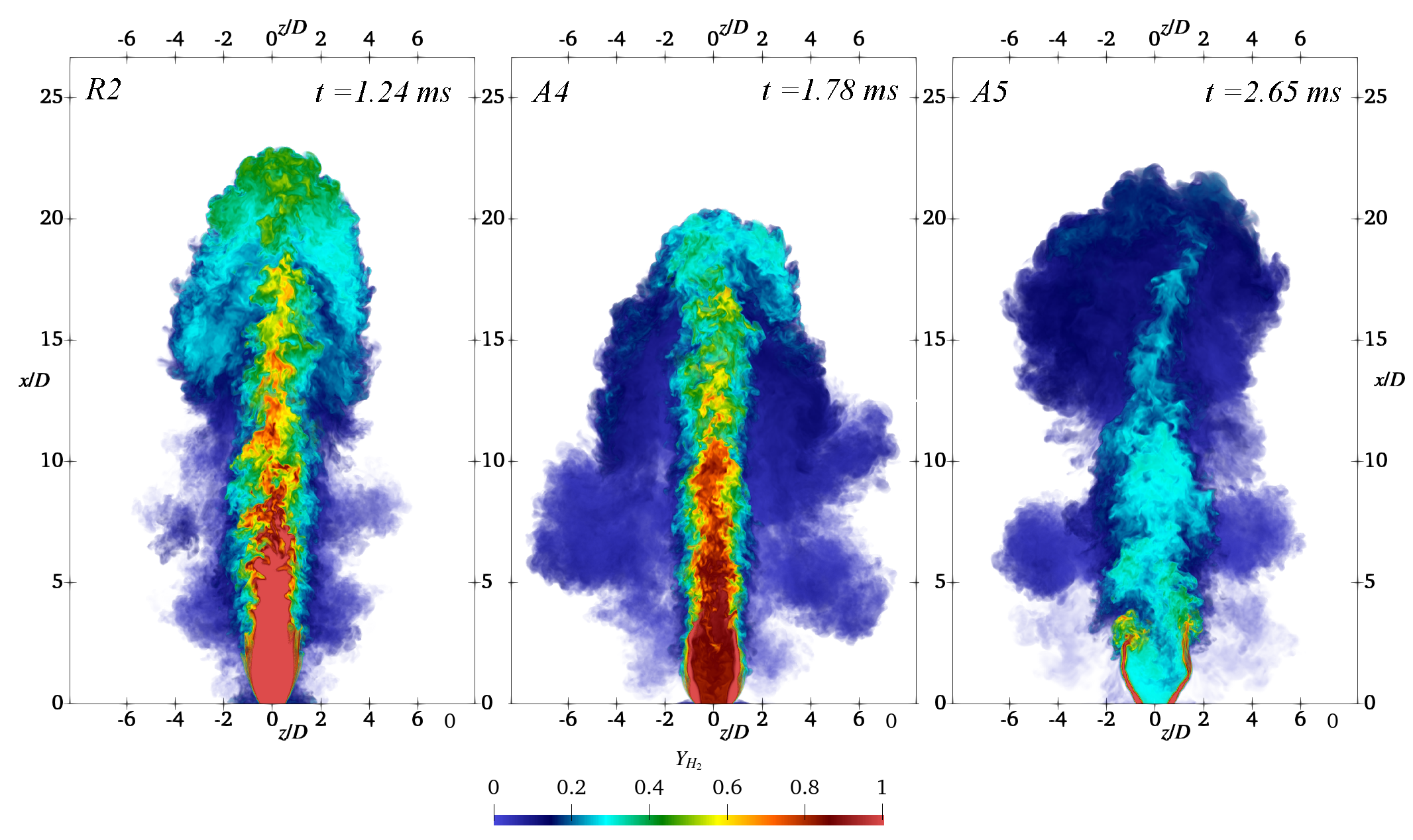}
\caption{Hydrogen mass fraction for the jets R2, A4 and, A5. Time is relative to a non-dimensional mass equal to 100 injected in the computational domain.}
\label{fig:yh2_mass}
\end{figure*}

The improvement obtainable with the hollow-cone configurations, particularly with the one in test case A5, is remarkable.  In the latter configuration, the high-concentration areas where the hydrogen is not flammable are almost entirely removed, and the average concentration decreases significantly. This is confirmed by figure \ref{fig:yh2_mass}, where we reported the hydrogen mass fraction at the time instant when the injected mass is the same.

For all the jets, we can observe a peak of the average  Y$_{H2}$ during the first phases of the injection. Almost all the hydrogen in the potential core remains within a few diameters downstream of the exit section,  and no mixing occurs despite the intense expansion.

The increased jet angle promotes the mixing with the surrounding air and almost completely removes the hydrogen-rich potential core. The region where the hydrogen mass fraction is close to one is shrunk, for case A5, to a hollow cone around the annular inflow section. 

\section{Conclusions}
We investigated hydrogen under-expanded jets using Large Eddy Simulations of round and annular nozzles, focusing on the turbulent mixing of the hydrogen jets to provide insights and valuable indications to optimise the mixture formation process in low-emission propulsion systems. 

We verified and validated the simulations against experimental particle image velocimetry and Schlieren images when available. 

The analyses cover different injection pressure ratios and nozzle geometrical configurations. 
The main outcomes are:
\begin{itemize}
    \item Transient evolution of hydrogen under-expanded jets radically differs from that observed for air. The mixing outside the jet core is enhanced by the baroclinic effect. The structure of the jet core is similar for the two flows in terms of shock structures.
    \item Hydrogen's physical properties trigger an intense interaction with the surrounding air. The hydrogen jet develops very high vorticity and engulfs a large air volume.
    \item  When increasing the pressure ratio, the repeated shock cells disappear, we have a unique Mach disc, and the potential core becomes smaller. The mixture quality slightly improves. 
    \item Hollow cone jets have a drastically different morphology when compared to the classical round jets. Ring-shaped shock cells substitute the Mach disc, the jet from highly under-expanded becomes weakly under-expanded, and injecting the same amount of hydrogen requires longer.
    \item The recirculation zone developing downstream of the nozzle for approximately 4D significantly enhances the mixing activity, removes the hydrogen-rich core, and gives a higher quality mixture.
    \item Large jet cone angles (i.e., 135$^\circ$) improve the mixture quality from the combustion point of view, quickly reducing the average hydrogen concentration already in the first phase of the injection.
\end{itemize}

\begin{acknowledgments}
This work has been funded by the European Union - NextGenerationEU, Mission 4, Component 1, under the Italian Ministry of University and Research (MUR) National Innovation Ecosystem grant ECS00000041 - VITALITY - CUP E13C22001060006.
\end{acknowledgments}

\section*{Data Availability Statement}

The data supporting this study's findings are available within the article.

\bibliography{aipsamp}

\end{document}